\documentclass[twocolumn,pra,aps,showpacs,superscriptaddress]{revtex4}

\usepackage{graphicx}
\usepackage{amssymb}
\usepackage{amsmath}
\usepackage[normalem]{ulem}
\begin{document}
\title{Coherent Backscattering of Light with Nonlinear Atomic Scatterers}

\author{T. Wellens}
\affiliation{Institut Non Lin{\'e}aire de Nice, UMR 6618, 1361 route
des Lucioles, F-06560 Valbonne}
\affiliation{Laboratoire Kastler Brossel, Universit{\'e} Pierre et Marie Curie,
4 Place Jussieu, F-75005 Paris}
\author{B. Gr{\'e}maud}
\author{D. Delande}
\affiliation{Laboratoire Kastler Brossel, Universit{\'e} Pierre et Marie Curie,
4 Place Jussieu, F-75005 Paris}
\author{C. Miniatura}
\affiliation{Institut Non Lin{\'e}aire de Nice, UMR 6618, 1361 route
des Lucioles, F-06560 Valbonne}

\date{\today}

\begin{abstract}
We study coherent backscattering of a monochromatic laser by a
dilute gas of cold two-level atoms in the weakly nonlinear regime. The
nonlinear response of the atoms results
in a modification of both the average field propagation (nonlinear
refractive index) and the scattering events. Using a perturbative
approach, the nonlinear effects arise from
inelastic two-photon scattering processes. We present a detailed
diagrammatic derivation of the elastic and
inelastic components of the backscattering signal both for scalar
and vectorial photons. Especially, we show that the
coherent backscattering phenomenon originates in some cases from the
interference between {\em three} different scattering amplitudes.
This is in marked contrast with the linear regime where it is due
to the interference between two different scattering amplitudes.
 In particular we show that, if elastically
scattered photons are filtered out from the photo-detection
signal, the nonlinear backscattering enhancement factor exceeds
the linear barrier two, consistently with a three-amplitude
interference effect.
\end{abstract}


\maketitle

\section{Introduction}

Propagation of light waves in disordered media is an active
research area since hundred years ago now. The original scientific
motivation came from astrophysical questions about properties of
light radiated by interstellar atmospheres \cite{schuster,
mihalas}. Then, within the first decades of the twentieth century,
the foundations of light transport in this regime were laid,
leading to the radiative transfer equations \cite{chandra, hulst,
holstein, molisch}. The basic physical ingredient of these
equations is a detailed analysis of energy transfers (scattering,
absorption, sources, \emph{etc}). Sufficiently far from any
boundaries, the long-time and large spatial scale limits of these
equations give rise, in the simplest cases, to a physically
appealing diffusion equation.

One important feature of this theory is to consider that any
possible interference effects are washed out under disorder
average. This is a \emph{random-phase} assumption. For a long
time, it was believed that this was still the case on average
for
monochromatic light elastically scattered off an optically thick
sample even if, for a given disorder realization, one observes a
speckle pattern \cite{dainty} indicating that phase coherence is
preserved by the scattering process. Theoretical and experimental
works in electronic transport \cite{anderson, langer, bergmann}
made soon clear that this random-phase assumption was wrong in the
elastic regime. Depending on the disorder strength, partial (weak
localization regime) or complete (strong localization regime)
suppression of diffusive behavior has been predicted, provided
phase coherence is preserved over a sufficient large number of
scattering events \cite{houches, akkermon}. In turn, these
discoveries have cross-fertilized the field of light transport in
the elastic regime \cite{photon, ucf, locaforte, dws}. In this
field, one of the hallmark of interference effects in elastic
transport is the \emph{coherent backscattering} (CBS) phenomenon
\cite{cbs, cbsdiffusion}: the average intensity multiply
scattered off an optically thick sample is larger than the average
background in a small angular range around the direction opposite
to the ingoing light. This interference enhancement of the diffuse
reflection off the sample is a manifestation of a two-wave
interference. As such, it probes the coherence properties of the
outgoing light and it has been extensively studied both
experimentally and theoretically. It can be shown on general
arguments, that the CBS enhancement factor (defined as the ratio
of the
backscattering CBS peak to diffuse background) never exceeds
the value 2 and is obtained in the helicity-preserving
polarization channel for scatterers with spherical symmetry
\cite{bvtmaynard}.

Whereas these interference modifications of transport are by now
widely understood in the case of linear media, recent experimental
developments have required an extension of multiple scattering
theory to the nonlinear case. Even if few studies already exists, they
only cover the simpler case of classical linear scatterers embedded in a
nonlinear medium~\cite{cbsnl1,cbsnl2}, whereas in our microscopic
approach, the nonlinear behavior of randomly distributed scatterers
will affect both the scattering processes and the average propagation.
In particular, with the
advent of laser cooling, on the one hand, it has become possible
to study interference effects in multiple scattering of light by
{\em cold atoms} \cite{labeyrie99, jonckheere00, cord1, cord2,
havey1}. In the regime where the saturation of the atomic
transition sets in, atoms scatter light nonlinearly, {\em i.e.}
the scattered light is no longer proportional to the incident one.
One should note that important nonlinear effects are \emph{easily}
achieved with atoms even at moderate laser intensities.
Considering a given driven optical dipole atomic transition, the
order of magnitude of the required light intensity to induce
nonlinear effects is given by the so-called saturation intensity
$I_s$ and is generally low. As typical examples, it is 1.6
mW/cm$^2$ for Rubidium atoms and 42 mW/cm$^2$ for Strontium atoms,
for their usual laser cooling transitions. On the other hand, {\em
random lasers} - mirrorless lasers where feedback is provided by
multiple scattering \cite{lethokov} - have been realized
experimentally \cite{lawandy94, cao99}. Here, nonlinear effects
occur in the regime close to or above the laser threshold. Since,
at least in the regime of coherent feedback \cite{cao},
interference is believed to play a decisive role in the physics
of the random laser, a better understanding of the influence of
nonlinearity (and amplification) on the properties of coherent
wave transport becomes necessary.

\section{Motivation and outline}

In a recent contribution \cite{wellens2}, we have shown that
nonlinear scattering may fundamentally affect interference in
multiple scattering. Indeed, in the perturbative regime of at most
one scattering event with $\chi^{(3)}$ nonlinearity, there are now
{\em three} (and no longer two) CBS interfering amplitudes.
Depending on the sign of the nonlinearity, {\em i.e.}
depending
whether nonlinear effects enhance or decrease the scattering cross
section, the effect of this \emph{three-wave} interference effect
leads to a significant increase or decrease of the nonlinear CBS
enhancement factor.

The purpose of the present paper is, on the one hand, to provide a
detailed derivation of the equations for the nonlinear coherent
backscattering signal used in \cite{wellens2}, and, on the other
one, to extend the treatment of \cite{wellens2} to the case of
atomic scatterers. Here, in contrast to the classical case, light
is scattered {\em inelastically}, {\em i.e.} the scattered photons
may change their frequencies. This leads to dephasing between
interfering amplitudes and, consequently, to a reduction of the
CBS enhancement factor in addition to the nonlinear modifications
mentioned above. Theoretical studies of this inelastic decoherence
mechanism have been so far restricted to the case of two atoms
\cite{wellens,slava,langevin}. Since the total (linear and
nonlinear) \emph{elastic} signal can be filtered out by
means of a suitable frequency-selective detection, a clear
experimental study of \emph{inelastic, nonlinear} CBS
becomes possible. Please note that this would be otherwise
very difficult to achieve since for weak intensities - the regime
where our theory is valid - the linear signal generally
largely dominates over the nonlinear one. In this paper, we will
show that the enhancement factor for inelastically scattered light
significantly exceeds the linear barrier two in certain frequency
windows. In contrast, the total enhancement factor - including
also elastically scattered light - is {\em diminished} by
nonlinear scattering. This is due to the negative sign of the
total nonlinear component, since the total (elastic plus
inelastic) scattering cross section is {\em decreased} by
saturation.

The paper is organized as follows. In Sec.~\ref{stheo}, we present
the perturbative theory for nonlinear CBS of light scattered off a
sample of cold two-level atoms. \lq Perturbative\rq\ here means
that we restrict ourselves to the regime of scalar - i.e., we
forget the polarization of the photon - two-photon
scattering with at most one nonlinear scattering event. This
assumption is valid at sufficiently low probe intensities and not
too large optical thicknesses. After shortly sketching the main
results of the linear case, Sec.~\ref{slin}, we derive equations
for the nonlinear backscattering signal in Sec.~\ref{snonl}. The
latter contains an inelastic and an elastic component. The latter
again splits into a nonlinear and a linear part. In
Sec.~\ref{pol}, supplemented by appendix \ref{appa}, we show how
to generalize our scalar theory to the vectorial case by
explicitly taking into account the light polarization degrees of
freedom. It is shown that nonlinear polarization effects lead to
decoherence between interfering paths. In contrast to the linear
case, this decoherence mechanism cannot be avoided by a suitable
choice of the polarization detection channel. In order to emphasize
the generality of our approach, we shortly discuss in Sec.~\ref{sclass}
a model of classical, nonlinear scatterers, which reproduces the elastic
backscattering signal of the atomic model. In
Sec.~\ref{results}, we apply our theory to the case of a
disordered atomic medium with slab geometry. We look at the
dependence of the backscattering signal as a function of the
optical thickness and of the detuning of the laser from the atomic
resonance. In particular, we show that the enhancement factor for
the inelastic component significantly exceeds the linear barrier
two in certain frequency windows. Finally, Sec.~\ref{concl}
concludes the paper.

\section{Theory}
\label{stheo}

In this section, we present the perturbative theory for
nonlinear coherent backscattering of light from a gas of cold two-level atoms.
We first treat the {\em linear} component of the backscattering
signal, which results from scattering of independent photons.
Thereby we introduce the reader, in Sec.~\ref{slin}, to standard
methods used in linear multiple scattering theory \cite{multiple},
which we will then generalize to the non-linear case in
Sec.~\ref{snonl}.

\subsection{Scalar linear regime}
\label{slin}

\subsubsection{One-photon scattering amplitude}

By definition, the linear component of the photo-detection signal
is proportional to the incoming intensity, in particular to the
number of photons in the initial laser mode. Since this implies
that the photons are independent from each other, it is sufficient
to know how a {\em single} photon propagates in the atomic medium,
see Fig.~\ref{lin}. This is equivalent to using the usual
Maxwell's equations for a disordered medium \cite{multiple}.

In the weak scattering regime, which we will consider throughout
this paper, transport is depicted as a succession of propagation
in an average medium interrupted by scattering events. The
important building block to properly describe scattering and
average propagation is the one-photon scattering amplitude by a
single atom. For near-resonant scattering, and for atoms with
\emph{no ground-state internal Zeeman degeneracies}, it reads~:
\begin{equation}
S_\omega=\frac{-4\pi i}{k(1-2i\delta/\Gamma)}.\label{scatt}
\end{equation}
It can be derived from the elastically-bound electron model in the
limit of small light detuning $\delta=\omega-\omega_{\rm
at}\ll\omega, \omega_{\rm at}$ \cite{multiple}. The atomic angular
transition frequency is $\omega_{\rm at}$ whereas the atomic
transition width $\Gamma$ describes radiative decay. The photon
wave number is $k$ and the photon angular frequency is $\omega=ck$
($c$ being the vacuum speed of light).

For simplicity, we work here with scalar photons, {\em i.e.} we
discard the vectorial nature of the light field. Scattering is
then fully isotropic and the differential scattering cross-section
simply reads
\begin{equation}
\frac{\mathrm{d}\sigma}{\mathrm{d}\Omega} =  \left|\frac{S_\omega}{4\pi}\right|^2 =
\frac{\sigma}{4\pi}
\end{equation}
leading to
\begin{equation}
\sigma = \frac{\sigma_0}{1+(2\delta/\Gamma)^2} \qquad ; \qquad
\sigma_0 = \frac{4\pi}{k^2}
\end{equation}
where $\sigma_0$ is the on-resonance scattering cross-section.

The scalar assumption is not a crucial one: as will be shown in
Sec.~\ref{pol}, the following treatment can be generalized to the
vectorial case. Please note however that the inclusion of internal
degeneracies is not immediately simple and requires a separate
treatment on its own. This is so because then the internal
dynamics is no longer simple (optical pumping sets in). \emph{In
this respect the results presented throughout this paper only
apply to non-degenerate ground-state atoms}. Please note also that
internal degeneracies are already known to strongly reduce the CBS
effect in the linear regime \cite{cord1,cord2}.

\subsubsection{Linear refraction index}

Between two successive scattering events occurring at ${\bf r}$ and
${\bf r'}$, the photon experiences an effective atomic medium with
refractive index $n_\omega$. Formally, the resulting propagation
is described by the average Green's function:
\begin{equation}
G_\omega({\bf r},{\bf r'})=- \frac{e^{i n_\omega k|{\bf r}-{\bf
r'}|}} {4\pi\,|{\bf r}-{\bf r'}|},\label{green}
\end{equation}
where the refractive index is given by \cite{index}:
\begin{equation}
n_\omega=1-\frac{\delta}{\Gamma k\ell}+\frac{i}{2k\ell}.\label{index}
\end{equation}
The imaginary part of $n_\omega$ describes depletion by
scattering. This depletion gives rise to the exponential
attenuation of the direct transmission through the sample
(Beer-Lambert law) and defines, \emph{via} the optical theorem,
the linear mean-free path at frequency $\omega$ as
\begin{equation}
\ell=\frac{1}{{\mathcal N}\sigma}
\label{ell}
\end{equation}
where $\mathcal N$ denotes the density number of atoms in the
sample. The weak scattering condition, where all the previous (and following)
results are valid, then simply reads $k\ell \gg 1$.

\subsubsection{Linear radiative transfer equation}
\begin{figure}
\centerline{\includegraphics[width=8cm]{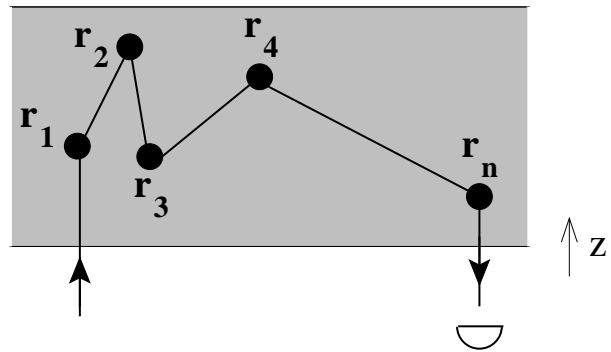}}
\caption{Scattering path of a single photon entering the medium
and leaving it in the backscattering direction to reach the
detector. Straight lines depict average propagation in the
effective medium while full circles depict scattering events
labeled by the $\textbf{r}_n$.\label{lin}}
\end{figure}
We have now at hand all the necessary ingredients to write down
the amplitude of a multiple scattering process as the one sketched
in Fig.~\ref{lin}. We consider a scattering volume $V$ exposed to
an initial monochromatic field with amplitude $E_0$ propagating
along axis $z$. The transverse area of the scattering volume is
$\Sigma$. Since $k\ell \gg 1$, a semi-classical picture using
well-defined scattering paths is appropriate. For a given
scattering path $\mathrm{C}_n \equiv ({\bf r_1} \to \dots \to {\bf
r_n})$ labeled by the collection of scattering events, the
corresponding far-field amplitude radiated at position
$\textbf{R}$ of the detector placed in the backscattering
direction is:
\begin{equation}
\mathcal{E}(\mathrm{C}_n) = - \frac{e^{ikR}}{4\pi kR}  \,
\mathcal{A}(\mathrm{C}_n) \, E_0.
\end{equation}

The complex amplitude $\mathcal{A}(\mathrm{C}_n)$ is simply a
product of one-photon scattering amplitudes (\ref{scatt}) and of
Green's functions (\ref{green}):
\begin{equation}
\mathcal{A}(\mathrm{C}_n) = kS_\omega \, e^{ikn_\omega (z_1+z_n)}
\left( \prod_{i=1}^{n-1} S_\omega G_\omega({\bf r_i},{\bf
r_{i+1}})\right)  \label{scattlin}
\end{equation}
where $z_i$ is the distance at which scattering event $i$ occurs
from the boundary of the medium. The superposition principle then
gives the total electric field amplitude $\mathcal{E}$ as a sum
over all possible scattering paths $\mathrm{C}_n$:
\begin{equation}
\mathcal{E} = - \frac{e^{ikR}}{4\pi kR} \, E_0 \, \mathcal{A}
\qquad ; \qquad \mathcal{A} = \sum_{\mathrm{C}_n}
\mathcal{A}(\mathrm{C}_n) \label{totalfield}
\end{equation}

The \emph{total average} intensity is obtained by squaring
(\ref{totalfield}) and averaging over all possible scattering
events. We define the total dimensionless bistatic coefficient as:
\begin{equation}
\gamma_{\rm el}^{(1)} = \frac{4\pi R^2}{\Sigma E_0^2} \,
\langle|\mathcal{E}|^2\rangle_{\mathrm{dis. av.}} = \frac{1}{4\pi
k^2 \Sigma} \, \langle|\mathcal{A}|^2\rangle_{\mathrm{dis. av.}}
\end{equation}

We now assume complete cancellation of interference effects
between different scattering paths (random-phase or Boltzmann
approximation). We then obtain the background (or \lq ladder\rq)
component of the backscattering signal:
\begin{equation}
\gamma_{\rm el}^{(1)} \approx L_{\rm el}^{(1)} = \sum_{n=1}^\infty
\frac{{\mathcal N}^n}{4\pi k^2 \Sigma} \int_V d{\bf r}_1 \dots
d{\bf r}_n \, |\mathcal{A}(\mathrm{C}_n)|^2 \label{iel1}
\end{equation}

This formula has a well-defined limit when $\Sigma \to \infty$ and
thus can be applied to slab geometries. Please note that, in
writing (\ref{iel1}), we have also discarded recurrent scattering
paths, {\em i.e.} paths visiting a given scatterer more than once.
Both approximations are justified in the case of a dilute medium,
$k\ell\gg 1$ \cite{recurr}.

We rewrite (\ref{iel1}) as
\begin{equation}
L_{\rm el}^{(1)} = \int \frac{d{\bf r}}{\Sigma\ell}\,
I_\omega({\bf r}) e^{-z/\ell}, \label{iel2}
\end{equation}
with
\begin{multline}
I_\omega({\bf r}) =  e^{-z/\ell}+\sum_{n=1}^\infty {\mathcal
  N}^n\int_V d{\bf r_1}\dots d{\bf r_n} \\
e^{-z_1/\ell}\prod_{i=1}^n\left|S_\omega G_\omega({\bf r_i},{\bf
    r_{i+1}})\right|^2,
\label{int1}
\end{multline}
where ${\bf r_{n+1}}={\bf r}$. This dimensionless function
describes the average light intensity at ${\bf r}$, in units of the
incident intensity $I_0=\epsilon_0 c E_0^2/2$ (in $W/m^2$) with
$\epsilon_0$ the vacuum permittivity. The first term in
Eq.~(\ref{int1}) represents the exponential attenuation of the
incident light mode, {\em i.e.} light which has penetrated to
position ${\bf r}$ without being scattered (Beer-Lambert law). The
remaining term describes the diffuse intensity, {\em i.e.} light
which has been scattered at least once before reaching ${\bf r}$.
From Eq.~(\ref{int1}), one can easily show that $I_\omega({\bf
r})$ fulfills the radiative transfer integral equation
\cite{chandra}:
\begin{equation}
I_\omega({\bf r})=e^{-z/\ell}+\frac{4\pi}{\ell}\int_V d{\bf r'}
|G_\omega({\bf r},{\bf r'})|^2 I_\omega({\bf r'})
\label{radtransf}
\end{equation}
The required solution of Eq.~(\ref{radtransf}) can be
obtained numerically by iteration starting from $I_\omega({\bf
r})=0$.

\subsubsection{Linear CBS cone}

In fact, the preceding Boltzmann approximation $\gamma_{\rm
el}^{(1)} \approx L_{\rm el}^{(1)}$ is wrong around the
backscattering direction. Indeed, on top of the background ladder
component, one observes a narrow cone of height $C_{\rm el}^{(1)}$
and angular width $\Delta\theta \propto (k\ell)^{-1}$
\cite{cbsdiffusion}. In the regime $k\ell \gg 1$, this so-called
CBS cone arises from the interference between amplitudes
associated to reversed scattering paths $\mathrm{C}_n \equiv ({\bf
r_1} \to \dots \to {\bf r_n})$ and $\widetilde{\mathrm{C}_n} \equiv
({\bf r_n} \to \dots \to {\bf r_1})$. Of course single scattering
paths where $n=1$ do not participate to this two-wave interference
(since they are exactly identical to their reversed counterparts)
and must be excluded from $C_{\rm el}^{(1)}$. Thereby, we obtain
the interference (or \lq crossed\rq) contribution as:
\begin{equation}
C_{\rm el}^{(1)} = \sum_{n=2}^\infty \frac{{\mathcal N}^n}{4\pi
k^2 \Sigma} \int_V d{\bf r}_1 \dots d{\bf r}_n \,
\mathcal{A}(\mathrm{C}_n) \,
\mathcal{A}^*(\widetilde{\mathrm{C}_n})\label{crossed2}
\end{equation}

Thus, the bistatic coefficient in the backscattering direction
reads $\gamma_{\rm el}^{(1)} = L_{\rm el}^{(1)}+C_{\rm el}^{(1)}$.
From Eq.~(\ref{scattlin}), we verify that the reciprocity symmetry
$\mathcal{A}(\mathrm{C}_n) =
\mathcal{A}(\widetilde{\mathrm{C}_n})$ is fulfilled for scatterers
without any internal ground-state degeneracies. This allows us to
rewrite (\ref{crossed2}) as
\begin{equation}
\begin{aligned}
C_{\rm el}^{(1)} &= \int \frac{d{\bf
r}}{\Sigma\ell}\left(I_\omega({\bf r})-e^{-z/\ell}\right) e^{-z/\ell} \\
&=  L_{\rm el}^{(1)} - S_{\rm el}^{(1)}
\label{crossed1}
\end{aligned}
\end{equation}
where $S_{\rm el}^{(1)}$ is the single scattering contribution.
Hence, the linear
CBS enhancement factor, defined as
\begin{equation}
\eta^{(1)}=1+C_{\rm el}^{(1)}/L_{\rm el}^{(1)} = 2 - S_{\rm
el}^{(1)}/L_{\rm el}^{(1)},
\end{equation}
is always smaller than two. It equals two if single scattering can
be filtered out, see Sec.~\ref{pol}.

\subsection{Scalar nonlinear regime}
\label{snonl}

At higher incident intensities, the successive photon scattering
events become correlated. Indeed absorption of one single
photon brings the atom in its excited state where it rests for a
quite long time $\Gamma^{-1}$ without being able to scatter other
incident photons. This means that saturation of the optical atomic
transition sets in, inducing nonlinear effects and inelastic
scattering. In a perturbative expansion of the photo-detection
signal in powers of the incident intensity, the leading nonlinear
term arises from scattering of {\em two} photons. In order to
generalize the above linear treatment to the two-photon case, we
first need to remind some relevant facts about scattering of two
photons by a single atom \cite{wellens}.

\subsubsection{One-atom two-photon inelastic spectrum}

The two-photon scattering matrix $S$ contains an elastic and
an inelastic part. The elastic part corresponds to two
single photons scattered independently from each other, whereas
the inelastic part describes a \lq true\rq\ two-photon scattering
process, where the photons become correlated and exchange energy
with each other. To obtain the intensity of the photo-detection
signal, the electric field operator $E$ (evaluated at the position
of the detector) is applied on the final two-photon state
$|f\rangle=S|i\rangle$, with $|i\rangle$ the initial state. Since
$E$ annihilates one photon, this yields a single-photon state
$|\psi\rangle=E|f\rangle$, which describes the final state of the
undetected photon. Like the scattering matrix $S$, it consists of
an elastic and an inelastic component:
\begin{equation}
|\psi\rangle=|\psi_{\rm el}\rangle+|\psi_{\rm in}\rangle.\label{psi}
\end{equation}
The inelastic part $|\psi_{\rm in}\rangle$ is a spherical wave emitted
by the atom, whereas the elastic part $|\psi_{\rm el}\rangle$ is a superposition of
scattered and unscattered light, thereby taking into account forward scattering of
the undetected photon. (Forward scattering of the detected photon does not need to be
taken into account, since the detector is placed in the backscattering direction.)
Finally,
the norm $I=\langle\psi|\psi\rangle$ of $|\psi\rangle$
defines the intensity of the photo-detection signal.
According to Eq.~(\ref{psi}), $I$ is the sum of the following three terms:
\begin{eqnarray}
I_{\rm el}^{(1)} & = & \langle\psi_{\rm el}|\psi_{\rm el}\rangle,\label{el1}\\
I_{\rm el}^{(2)} & = & 2{\rm Re}\{\langle \psi_{\rm el}|\psi_{\rm
in}\rangle\},\label{el2}\\
I_{\rm in}^{(2)} & = & \langle\psi_{\rm in}|\psi_{\rm in}\rangle.\label{in}
\end{eqnarray}
So far, everything is valid for any two-photon scattering process with an elastic and
an inelastic component. In the specific case of a single atom, the following
result is obtained:
\begin{eqnarray}
I_{\rm el}^{(1)} & = & \frac{\sigma}{4\pi R^2} \, I_0 \label{el12}\\
I_{\rm el}^{(2)} & = & -2 I_{\rm el}^{(1)}s,\label{el22}\\
I_{\rm in}^{(2)} & = & I_{\rm el}^{(1)} s \label{in2},
\end{eqnarray}
with the incident intensity $I_0$, and the saturation parameter $s$
defined by \cite{cct}:
\begin{equation}
s = \frac{s_0}{1+(2\delta/\Gamma)^2}, \quad s_0 = \frac{I_0}{I_s},
\quad I_s = \epsilon_0 c \left(\frac{\hbar \Gamma}{2d}\right)^2
\end{equation}
where $d$ is the atomic dipole strength and $I_s$ the saturation
intensity of the atomic transition.

The first term, Eqs.~(\ref{el1},\ref{el12}), which arises from two
photons scattered independently from each other, reproduces the
linear single-photon cross section $4\pi\sigma=|S_\omega|^2$, see
Eq.~(\ref{scatt}). The following two terms correspond to nonlinear
elastic and inelastic scattering, respectively. For the case of a
single atom, the perturbative two-photon treatment is valid for
$s\ll 1$, {\em i.e.} if the nonlinear terms are small compared to
the linear one.

The frequency spectrum of the elastically scattered light is
simply $F_{\rm el}(\omega') = (I_{\rm el}^{(1)}+I_{\rm
el}^{(2)})\,\delta(\omega'-\omega)$ whereas the frequency spectrum
of the inelastically scattered light is $F_{\rm in}(\omega') = I_{\rm
in}^{(2)} \, P(\omega')$. The continuous spectrum $P(\omega')$ is
normalized to unity according to $\int d\omega' P(\omega') =1$. It
is obtained as follows \cite{wellens}:
\begin{equation}
P(\omega') = \frac{\Gamma}{4\pi}
\left|\frac{1}{\delta'+i\Gamma/2}+
\frac{1}{2\delta-\delta'+i\Gamma/2}\right|^2\label{specin},
\end{equation}
where $\delta'=\omega'-\omega_{\rm at}$ denotes the final
detuning. This inelastic spectrum consists of two peaks with width
$\Gamma$, one located at the atomic resonance
($\omega'=\omega_{\rm at}$), and the other one twice as far
detuned as the incident laser ($\omega'=\omega_{\rm at}+2\delta$).  For
$\delta<\Gamma/2$, the two peaks
merge to a single one centered at $\omega'=\omega$.
Please note that, by going beyond the
two-photons scattering approximation, one would then get
\emph{three} peaks as predicted by the nonperturbative calculation
of the inelastic spectrum, also known as the Mollow triplet
\cite{cct}.

\subsubsection{Nonlinear scattering in a dilute medium of atoms}

\begin{figure}
\centerline{\includegraphics[width=8cm]{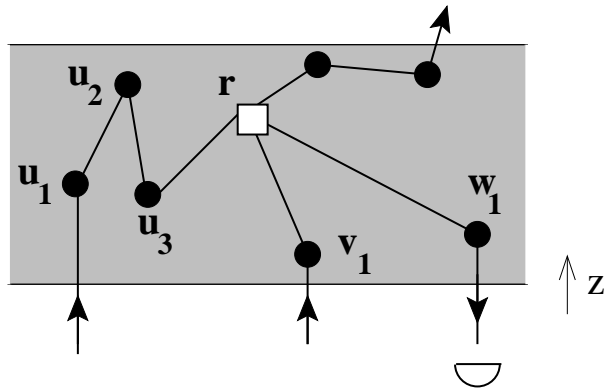}} \caption{In
the perturbative approach, we assume a single nonlinear two-photon
scattering event ($\Box$), but arbitrarily many linear scattering
events ($\bullet$). One of the two photons is finally annihilated
by the detector, thereby defining the photo-detection signal,
whereas the other one is scattered into an arbitrary
direction.\label{nonl1}}
\end{figure}
Now, we generalize the above single-atom treatment to a multiple
scattering process in a dilute medium of atoms. First, we note
that the above perturbative treatment - in particular
Eqs.~(\ref{el1}-\ref{in}) - remains valid for any form of the
scattering sample, be it a single atom, two atoms, or arbitrarily
many of them. An important difference from the single-atom case,
however, is that the total weight of nonlinear processes may be
drastically enhanced if the sample has a large optical thickness
$b=L/\ell$, where $L$ is the typical medium size. This implies
that the condition $s\ll 1$ is not sufficient to guarantee the
validity of the perturbative approach. Instead, as we will argue
in Sec.~\ref{results}, the perturbative condition reads $sb^2\ll
1$.

A typical two-photon scattering path is sketched in
Fig.~\ref{nonl1}. Here, the incoming photons propagate at first
independently from each other to position ${\bf r}$
inside the
disordered atomic medium, where they undergo a nonlinear
scattering event. One of the two outgoing photons then propagates
back to the detector. The possibility that the two photons  meet
again at another atom can be neglected in the case of a dilute
medium, similar to recurrent scattering in the linear case
\cite{recurr}. We can hence restrict our analysis to processes
like the one shown in Fig.~\ref{nonl1}, with arbitrary numbers of
linear scattering events before and after the nonlinear one. Thus
one of the two incoming photons undergoes $n\geq 0$ \emph{elastic}
scattering events (labeled by ${\bf u_i}$), while the other
undergoes $m\geq 0$ \emph{elastic} scattering events (labeled by
${\bf v_j}$), before merging at ${\bf r}$ where they undergo the
\emph{inelastic} scattering event. One of the outgoing inelastic
photons reaches back the detector after having undergone $l\geq 0$
\emph{elastic} scattering events (labeled by positions ${\bf
w_k}$). For the other \emph{undetected} inelastic photon, we may
assume, without any loss of generality, that it does not interact
anymore with the atomic medium. This interaction would be anyway
described by a unitary operator (as a consequence of energy
conservation), which does not change the norm of the state
$|\psi\rangle$ of the undetected photon defining the detection
signal.

In general, the state of the \emph{inelastic} undetected photon
corresponding to a scattering path $\mathrm{C}$ defined by the
position ${\bf r}$ of the {\em two-photon} scattering event and by
the collection of positions of all \emph{one-photon} scattering
events $\mathrm{C} \equiv \{{\bf u},{\bf v},{\bf r},{\bf w}\}$ is
given as follows:
\begin{multline}
|\psi_{\rm in}(\mathrm{C})\rangle = e^{i kn_\omega (z_{u_1}+z_{v_1})}\\
\times\prod_{i=1}^{n}S_\omega G_\omega({\bf u_i},{\bf u_{i+1}})
\prod_{j=1}^{m}S_\omega G_\omega({\bf v_j},{\bf v_{j+1}})\\
\times\int d\omega'\,\Pi_{\omega'}|\psi_{\rm in}\rangle
\prod_{k=1}^{l}S_{\omega'} G_{\omega'}({\bf w_k},{\bf w_{k+1}})\\
e^{i kn_{\omega'} z_{w_1}}\times
\begin{cases}
1 & n=m=0,\\
 2 & n>0\text{ or } m>0
\end{cases}\label{snonlin}
\end{multline}
with ${\bf u_{n+1}}={\bf v_{m+1}}={\bf w_{l+1}}={\bf r}$,
$\Pi_{\omega'}$ the projector on photons states at frequency
$\omega'$ and $|\psi_{\rm in}\rangle$ the inelastic final state of
the \emph{one-atom} case, Eq.~(\ref{psi}). Since the inelastic
two-photon scattering event takes place at position ${\bf r}$, this
state describes an outgoing spherical wave emitted at ${\bf r}$.
Furthermore, note that if the two incoming photons do not
originate both from the incident mode, \emph{i.e.} if $n>0$ or
$m>0$, a factor $2$ arises due to the fact that the incoming
photons can be distributed in two different ways among the paths
$\{u\}$ and $\{v\}$.

\begin{figure}
\centerline{\includegraphics[width=8cm]{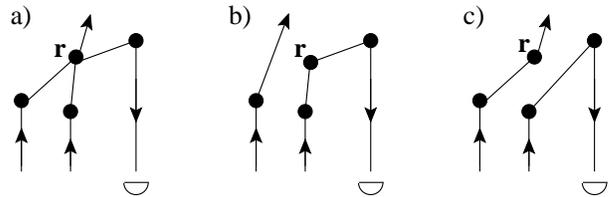}} \caption{The
elastic component $|\psi({\bf r},\{{\bf u},{\bf v},{\bf w}\})_{\rm
el}\rangle$ of the undetected photon's state arises from a
superposition of the following three processes: (a) elastic
scattering of both photons at ${\bf r}$, (b) only the detected
photon scattered at ${\bf r}$, and (c) only the undetected photon
scattered at ${\bf r}$. The last two diagrams are necessary to take
into account the nonlinear average propagation of the undetected (b) or detected
(c) photon. \label{figel}}
\end{figure}
The {\em elastic} component $|\psi_{\rm el}(\mathrm{C})\rangle$ is
obtained in a similar way. However, as in the single-atom case, we
must take into account forward scattering of the undetected
photon, at the position ${\bf r}$ of the nonlinear event. This is
done by considering the superposition of two diagrams where the
undetected photon is scattered or not scattered at ${\bf r}$, see
Fig.~\ref{figel}(a,b). Since this approach exactly parallels the
one known from the single-atom case \cite{wellens}, it is
unnecessary to present the complete calculation of the elastic
component in detail - all relevant ingredients to perform the
generalization to the multi-atom case will be contained in the
calculation of the inelastic component. In contrast to the
single-atom case, however, the elastic component will enter in the
calculation of the {\em nonlinear average propagation}, {\em i.e.} the nonlinear
modification of the refractive index (Kerr effect), and will be
discussed later. At first, we concentrate on the
processes of {\em nonlinear scattering}, {\em i.e.} processes
changing the direction of propagation of the detected photon.

As for the linear case, we still assume the same dilute medium
approximations to hold for the \lq ladder\rq\ and \lq crossed\rq\
contributions. Thus, in order to calculate the average
photo-detection signal, we just keep scattering diagrams obtained
by reversing the path of the detected photon. Furthermore, we also
neglect interference between diagrams where the nonlinear
scattering event occurs at different atoms. This is justified in
the dilute case since the overlap between two spherical waves
emitted at ${\bf r}$ and ${\bf r'}$ vanishes if $k|{\bf
r}-{\bf r'}|\gg 1$.

\subsubsection{Nonlinear ladder contribution}\label{snonlladder}

To obtain the inelastic component of the average backscattering
signal, we first get the total final state of the undetected
photon by summing Eq.~(\ref{snonlin}) over all possible different
scattering paths $\mathrm{C}$. Then we insert this result into
Eq.~(\ref{in}) and we finally average over the random positions of
the scatterers. As argued above, only identical or reversed
scattering paths are retained in the average, giving rise to the
background (\lq ladder\rq) and interference (\lq crossed\rq)
component. Thus, the inelastic background component reads as
follows:
\begin{multline}
L_{\rm in}^{(2)} =  \int_V d{\bf
r}\sum_{(n,m,l)=0}^\infty\frac{{\mathcal N}^{n+m+l+1}}{4\pi
k^2\Sigma}\int_V
\prod_{i=1}^n d{\bf u_i}\prod_{j=1}^m d{\bf v_j} \prod_{k=1}^l d{\bf w_k}\\
\langle \psi_{\rm in}(\mathrm{C})|\psi_{\rm
in}(\mathrm{C})\rangle \times
\begin{cases}
1 & \text{ if } n=m=0 \\
1/2 & \text{ otherwise}
\end{cases}\label{lnonlin}
\end{multline}

Note that some care must be taken not to sum twice over the same
scattering path. In particular, any exchange of the two incoming
parts $\{u\}$ and $\{v\}$ leaves the total scattering path
unchanged since the two incoming photons are identical. For this
reason, a factor $1/2$ must be inserted at the end of
Eq.~(\ref{lnonlin}). Again, as in Eq.~(\ref{snonlin}), the case
$n=m=0$ is exceptional, since then there is \emph{no} elastic
scattering events before the nonlinear one: the two incident
photons remain in the same mode.

If we insert now Eq.~(\ref{snonlin}) into Eq.~(\ref{lnonlin}), we
simply obtain the inelastic nonlinear ladder contribution as:
\begin{equation}
L_{\rm in}^{(2)}  =  s \int \frac{d{\bf r}}{\Sigma\ell}
\left(2I^2_\omega({\bf r})-e^{-2z/\ell}\right) \int d\omega'
P(\omega')I_{\omega'}({\bf r}),\label{nonlinl2}
\end{equation}
with $I_\omega({\bf r})$ the linear average intensity, see Eq.~(\ref{radtransf}).
In order to interpret this result, we first note that the
inelastic intensity radiated by the atom at position ${\bf r}$ is
proportional to the mean {\em squared} intensity at ${\bf r}$. An
alternative, physically transparent derivation of the latter can
be performed as follows: we write the local field amplitude
$\mathcal{A}=\exp(-z/2\ell)+\mathcal{A}_D$ as a sum of coherent
and diffuse light amplitudes. The latter term exhibits a Gaussian
speckle statistics \cite{goodman}, {\em i.e.} $\langle
\mathrm{Re}\mathcal{A}_D\rangle=\langle
\mathrm{Im}\mathcal{A}_D\rangle=0$, $2\langle
(\mathrm{Re}\mathcal{A}_D)^2\rangle=2\langle(\mathrm{Im}\mathcal{A}_D)^2\rangle=\langle
|\mathcal{A}_D)|^2\rangle$ and
$\left<|\mathcal{A}_D|^4\right>=2\left<|\mathcal{A}_D|^2\right>^2$.
Thereby, we obtain for the mean squared intensity:
\begin{eqnarray}
\left<|\mathcal{A}|^4\right> & =
&e^{-2z/\ell}+\left<|\mathcal{A}_D|^4\right>+4e^{-z/\ell}\left<|\mathcal{A}_D|^2\right>\label{fluc1}\\
& = & 2 \left<|\mathcal{A}|^2\right>^2-e^{-2z/\ell}.\label{fluc2}
\end{eqnarray}
Inserting the average intensity,
$I_\omega=\left<|\mathcal{A}|^2\right>$, we immediately recognize
the first integrand in Eq.~(\ref{nonlinl2}). Then, the atom emits
a photon with frequency distribution $P(\omega')$. Finally, due to
time reversal symmetry, the propagation of this photon from ${\bf
r}$ to the detector is described by the same function
$I_{\omega'}({\bf r})$ which represents propagation of incoming
photons to ${\bf r}$.

Concerning the elastic component, the diagrammatic calculation
\emph{via} Eq.~(\ref{el2}), see also Fig.~\ref{figel}(a,b), shows that
the above argument can be repeated in the same way - except for
the fact that the detected photon does not change its frequency.
Furthermore, a factor $-2$ is taken over from the single-atom
expression, cf. Eqs.~(\ref{el22},\ref{in2}). Thereby, we obtain:
\begin{equation}
L_{\rm el}^{(2,{\rm scatt})}  =  -2s \int \frac{d{\bf
r}}{\Sigma\ell} \left(2I^2_\omega({\bf
r})-e^{-2z/\ell}\right)I_{\omega}({\bf r}).\label{lel2scatt}
\end{equation}
The index \lq scatt\rq\ reminds us that we have treated only
nonlinear scattering so far.
Below (Sec.~\ref{sprop}), we will add nonlinear average
propagation, which contributes to the elastic nonlinear component, too.

\begin{figure}
\centerline{\includegraphics[width=7.5cm]{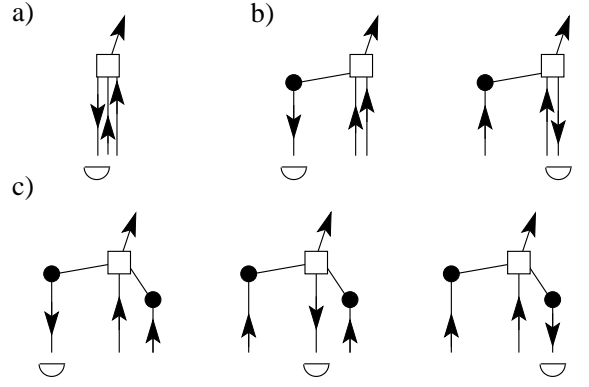}}
\caption{In the presence of nonlinear scattering ($\Box$),
there may be either (b) two, or (c) three
interfering amplitudes contributing to enhanced backscattering,
apart from single scattering (a), which only contributes to the background.
In general,
the case (c), which corresponds to maximum enhancement factor three,
is realized if either both incoming photons, or one incoming and the
outgoing detected photon exhibit at least
one linear scattering event ($\bullet$) besides the nonlinear one.
\label{fig1}}
\end{figure}

\subsubsection{Nonlinear crossed contribution}\label{snonlcrossed}

It remains to calculate the \lq crossed\rq\ contribution,
\emph{i.e.} interference between reversed paths. In contrast to
the linear case, where there are always {\em two} interfering
amplitudes (apart from single scattering), the nonlinear case
admits more possibilities to reverse the path of the detected
photon. This is due to the photon exchange symmetry at the
nonlinear scattering event, which does not allow to distinguish
which one of the two incoming photons finally corresponds to the
detected or undetected one. As evident from Fig.~\ref{fig1}(c),
each multiple scattering path where both incoming photons, or one
incoming and the outgoing detected photon, exhibit at least one
linear scattering event besides the nonlinear one, has {\em two}
different reversed counterparts, leading in total to {\em three}
interfering amplitudes.

If we look at the scattering process shown in Fig.~\ref{nonl1},
the two reversed counterparts are obtained by exchanging the
outgoing detected photon $\{w\}$ with either one of the incoming
photons $\{u\}$ or $\{v\}$. Since both cases are identical in the
ensemble average, we may restrict ourselves to one of them, let us
say $\{v\}$. We thus note $\widetilde{\mathrm{C}}\equiv \{{\bf
u},{\bf w},{\bf r},{\bf v}\}$ the reverse path corresponding to
$\mathrm{C} \equiv \{{\bf u},{\bf v},{\bf r},{\bf w}\}$ when
$\{v\}$ and $\{w\}$ are exchanged. In total, we obtain for the
inelastic interference component:
\begin{multline}
C_{\rm in}^{(2)} = \int_V d{\bf
r}\sum_{(n,m,l)=0}^\infty\frac{{\mathcal N}^{n+m+l+1}}{4\pi
k^2\Sigma}\times\\
\int_V\prod_{i=1}^n d{\bf u_i}\prod_{j=1}^m d{\bf v_j}
\prod_{k=1}^l d{\bf w_k} \langle \psi_{\rm
in}(\mathrm{C})|\psi_{\rm in}(\widetilde{\mathrm{C}})\rangle
\\\times
\begin{cases}
0 & \text{ if } m=l=0 \\
1 & \text{ otherwise}
\end{cases}\label{nonlinc}
\end{multline}
Here, the case $m=l=0$ identifies processes where
the two reversed paths $\mathrm{C}$ and $\widetilde{\mathrm{C}}$ are indistinguishable.
Setting their contribution equal to zero accounts in particular for the single
scattering
case depicted in Fig.~\ref{fig1}(a), {\em i.e.} $n=m=l=0$, which does not
contribute to the interference cone. The case Fig.~\ref{fig1}(b) remains with
{\em two} contributions ($n=m=0$, $l>0$, and $n=l=0$, $m>0$,
respectively) in Eq.~(\ref{nonlinc}), corresponding to the fact that
two amplitudes interfere. Finally, the case (c) of three interfering
amplitudes is reflected in Eq.~(\ref{nonlinc}) by the absence of the
exchange factor $1/2$, as compared to the background, Eq.~(\ref{lnonlin}).
Thereby, the interference contribution can, in principle,
become up to two times larger than the background.

If we insert the state of the undetected
photon, Eq.~(\ref{snonlin}), into Eq.~(\ref{nonlinc}), we encounter the following
expression
\begin{multline}
g_{\omega,\omega'}({\bf r}) = 
e^{ik(n_\omega-n_{\omega'}^*)z}+\sum_{n=1}^\infty{\mathcal N}^n
\int_V d{\bf r_1}\dots d{\bf r_n}\\
e^{ik(n_\omega-n_{\omega'}^*)z_1}\prod_{i=1}^n
S_\omega G_\omega({\bf r_i},{\bf
  r_{i+1}})S_{\omega'}^*G_{\omega'}^*({\bf r_i},{\bf r_{i+1}})
\label{radtransfc},
\end{multline}
which generalizes the local intensity, Eq.~(\ref{int1}), to the case where two
different frequencies
occur in the interfering paths.
Numerically, it
can be obtained as the iterative solution of:
\begin{multline}
g_{\omega,\omega'}({\bf r}) =  e^{ik(n_\omega-n_{\omega'}^*)z}+{\mathcal N}
S_\omega S_{\omega'}^*\\
\times\int_V d{\bf r'}G_\omega({\bf r},{\bf r'})G_{\omega'}^*({\bf r},{\bf r'})
g_{\omega,\omega'}({\bf r'}).
\label{radtransfc2}
\end{multline}
This function describes the ensemble-averaged product of two
probability amplitudes, one representing an incoming photon with
frequency $\omega$ propagating to position ${\bf r}$, and the other
one the complex conjugate of a photon with frequency $\omega'$
propagating from ${\bf r}$ to the detector. If
$\omega\neq\omega'$, then these amplitudes display a nonvanishing
phase difference both due to scattering and to average propagation
in the medium. This leads on average to a decoherence mechanism
and consequently to a loss of interference contrast. Indeed, both
the complex scattering amplitude, Eq.~(\ref{scatt}), and the
refractive index, Eq.~(\ref{index}), depend on frequency. In
contrast, the phase difference due to free propagation ({\em i.e.}
in the vacuum) can be neglected if $\Gamma\ell\ll c$, which is
fulfilled for typical experimental parameters \cite{thierry,
havey2}. In the case $\omega=\omega'$ of identical frequencies,
$g_{\omega,\omega}({\bf r})=I_\omega({\bf r})$ reduces to the
average intensity, see Eq.~(\ref{radtransf}).

In terms of the iterative solution of
Eq.~(\ref{radtransfc}), the inelastic interference term,
Eq.~(\ref{nonlinc}), is rewritten as follows:
\begin{equation}
\begin{aligned}
C_{\rm in}^{(2)}  = & 4 s \int d\omega'P(\omega')\int_V
\frac{d{\bf r}}{\Sigma\ell} \Bigl[I_\omega({\bf r})
|g_{\omega,\omega'}({\bf r})|^2\Bigr.\\
& -e^{-z/\ell} {\rm Re}\left\{e^{i(n_\omega-n_{\omega'}^*)kz}
g_{\omega,\omega'}^*({\bf r})\right\}\\
& \Bigl.-\left(I_\omega({\bf r})-e^{-z/\ell}\right)e^{-z/\ell-z/\ell'}\Bigr],
\label{nonlinc2}
\end{aligned}
\end{equation}
with $\ell'$ the linear mean free path at frequency $\omega'$.
In the elastic case, where $\omega'=\omega$,
dephasing between reversed scattering paths does not occur, and
the expression (\ref{nonlinc2}) simplifies to:
\begin{equation}
C_{\rm el}^{(2,{\rm scatt})}= -8 s\int_V \frac{d{\bf
r}}{\Sigma\ell} \left(I_\omega({\bf r})^3- 2I_\omega({\bf
r})e^{-2z/\ell}+e^{-3z/\ell}\right).\label{cel2scatt}
\end{equation}
Since, in the elastic case, there is no loss of coherence due to
change of frequency, the elastic interference component,
Eq.~(\ref{cel2scatt}), is completely determined by the relative
weights of the one-, two-, and three-amplitudes cases exemplified
in Fig.~\ref{fig1}. This can be checked by rewriting the
background and interference components,
Eqs.~(\ref{lel2scatt},\ref{cel2scatt}), in terms of diffuse and
coherent light, respectively, \emph{i.e.} by writing
$I=I_D+\exp(-z/\ell)$. One obtains:
\begin{equation}
\begin{aligned}
L_{\rm el}^{(2,{\rm scatt})} & \propto  \bigl<e^{-3 z/\ell}+5I_De^{-2z/\ell}
+~6I_D^2e^{-z/\ell}+2I_D^3\bigr>,\\
C_{\rm el}^{(2,{\rm scatt})} & \propto \bigl<
\underbrace{\phantom{e^{-3z/\ell}+\!}}_{\displaystyle
(a)}~\underbrace{4I_De^{-2z/\ell}}_{\displaystyle
(b)}+\underbrace{12I_D^2e^{-z/\ell}+4I_D^3}_{\displaystyle (c)}\bigr>,
\label{laddcross}
\end{aligned}
\end{equation}
where the brackets denote the integral over the volume $V$ of the medium, and
(a,b,c) correspond to the three cases
shown in Fig.~\ref{fig1}, identified by different powers of diffuse or
coherent light. As expected, the three-amplitudes case (c) implies an
interference term twice as large as the background.
In the two-amplitudes case (b),
a small complication arises, since one of the two interfering amplitudes is twice
as large as the other one ({\em i.e.} the one where
both incoming photons originate from the coherent mode),
cf. the discussion after Eq.~(\ref{snonlin}). In this case,
the interference contribution, $2\times 1+1\times 2=4$, is smaller than the background,
$2\times 2+1\times 1=5$.
Finally, as it should be, the single scattering term (a) is absent in the
interference term
$C_{\rm el}^{(2,{\rm scatt})}$.

\subsubsection{Nonlinear average propagation}
\label{sprop}

\begin{figure}
\centerline{\includegraphics[width=8.5cm]{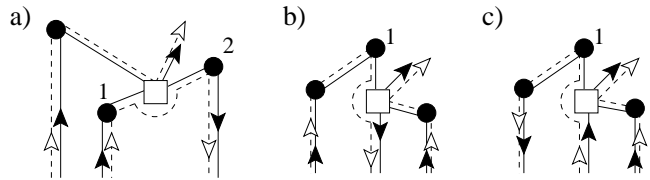}}
\caption{Diagrammatic description of nonlinear
propagation. A
two-photon process (solid lines)
interferes with two independent single photons
(dashed lines). Only one of the latter (the undetected photon)
is scattered at $\Box$, thereby modifying the propagation of the
detected photon
(a) between two scattering events at positions 1 and 2,
(b) on the way to the detector after the last scattering event at position 1,
or (c) in the coherent mode before the first scattering event at position 1.
\label{kerr}}
\end{figure}
So far, we have only considered processes of nonlinear scattering where
the direction of propagation of the detected photon is changed.
It remains to take into account nonlinear average propagation. This is
described by
those processes where, in one of the two interfering amplitudes,
the detected photon is not scattered at the position ${\bf r}$ of the
nonlinear event
\footnote{In principle, there exists also the possibility
that both the detected and undetected photon are not scattered at ${\bf r}$ in
one of the interfering amplitudes, corresponding to forward scattering
of both photons by the same atom. It can be shown, however, that this
case is negligible in the case of a dilute medium.}.
The corresponding diagrams
are depicted in Fig.~\ref{kerr}, where the two interfering amplitudes are
represented by the solid and dashed lines, respectively.
Here, the solid lines correspond to an inelastic two-photon scattering process
(like the one shown in Fig.~\ref{nonl1}), whereas the dashed lines represent
an elastic process, where the two photons are independent from each other,
see Fig.~\ref{figel}(c). Hence, their interference contributes to the
nonlinear {\em elastic} component of the photo-detection signal, cf. Eq.~(\ref{el2}).

The three diagrams shown in Fig.~\ref{kerr} differ only by the
fact that the nonlinear propagation event takes place either
between two scattering events at position 1 and 2 (a), on the way
to the detector, {\em i.e.} after the last scattering event at
position 1 (b), or in the coherent mode, {\em i.e.} before the
first scattering event at position 1 (c). At first, let us examine
the case (a). We imagine that each of the three dots $\bullet$ may
represent an arbitrary number of scattering events. [Only note
that the number of events corresponding to the dots and 1 and 2
must be larger than zero - otherwise, the diagram
Fig.~\ref{kerr}(a)  would be identical to Fig.~\ref{kerr}(b) or
(c).] According to the theory of linear radiative transfer
outlined in Sec.~\ref{slin}, the ladder diagrams corresponding to
the two incoming photons arriving at 1 (position ${\bf r_1}$) and at
the nonlinear event $\Box$ (position ${\bf r_3}$) yield the linear
local intensities $I_\omega({\bf r_1})$ and $I_\omega({\bf r_3})$,
respectively. Likewise (due to reciprocity symmetry), the
propagation of the outgoing detected photon from 2 (position ${\bf
r_2}$) to the detector - with arbitrary number of scattering
events in between - is given by $I_\omega({\bf r_2})$. Hence, the
only ingredient which we have to calculate is the nonlinear
propagation between 1 and 2. Note that, when taking the average
over the position ${\bf r_3}$ of the nonlinear event, non-negligible
contributions arise only if ${\bf r_3}$ is situated on the straight
line between ${\bf r_1}$ and ${\bf r_2}$, since this is the only way
to fulfill a stationary phase (or phase matching)
condition. Thereby, the \lq pump intensity\rq\ entering in the
nonlinear propagation is given by the average value of the local
intensity on this line, which we denote by $\langle
I_\omega\rangle_{{\bf
    r_1}\to{\bf r_2}}$.
We do not want to present the complete calculation here
(this requires to
calculate at first the case of a single atom, which can be done
with the techniques described in \cite{wellens}), but just give
the final result:
\begin{equation}
\left|G_\omega^{(\rm nl,a)}({\bf r_1},{\bf r_2})\right|^2 = \left|G_\omega({\bf
r_1},{\bf r_2})\right|^2 \frac{2sr_{12}}{\ell}
\langle I_\omega\rangle_{{\bf r_1}\to{\bf r_2}}.\label{propnl}
\end{equation}
From this, we deduce the following value for the nonlinear mean
free path:
\begin{equation}
\frac{1}{\ell^{(\rm nl)}({\bf r})}=\frac{1}{\ell}\left(1-2sI_\omega({\bf r})\right),
\label{pathd}
\end{equation}
which is consistent with Eq.~(\ref{propnl}), if we expand the
resulting propagator (where the mean free path appears in the exponent)
up to first order in $s$. The same result is also obtained in the case
of diagram Fig.~\ref{kerr}(b), {\em i.e.} for the propagation after the last
scattering event. Hence, the corresponding propagator (first order in $s$) reads:
\begin{equation}
\left|G_\omega^{(\rm nl,b)}({\bf r_1})\right|^2=e^{-z_1/\ell}
\frac{2s z_1}{\ell}
\langle I_\omega\rangle_{{\bf r_1}\to{\bf r_0}},\label{propnlc}
\end{equation}
where ${\bf r_0}={\bf r_1}-z_1 {\bf e_z}$, with ${\bf e_z}$ the unit vector
pointing in the direction of the incident laser, denotes the point
where the photon leaves the medium.
In the case (c), a small complication
arises since the photons arriving at the nonlinear event $\Box$ may originate
both from the coherent mode, which reduces the
two-photon scattering amplitude by a factor $1/2$, cf. the discussion after
Eq.~(\ref{snonlin}).
Hence, the nonlinear mean free path for photons from the coherent mode reads:
\begin{equation}
\frac{1}{\ell_c^{(\rm nl)}({\bf r})}=\frac{1}{\ell}\left(1-2sI_\omega({\bf r})+
s e^{-z/\ell}\right),\label{pathc}
\end{equation}
with the corresponding propagator
\begin{equation}
\left|G_\omega^{(\rm nl,c)}({\bf r_1})\right|^2=e^{-z_1/\ell}
\frac{s z_1}{\ell}
\left< 2I_\omega({\bf r})-e^{-z/\ell}\right>_{{\bf r_0}\to{\bf r_1}},\label{propnlb}
\end{equation}
The difference between the mean free paths, Eqs.~(\ref{pathd},\ref{pathc}),
can also be understood as a consequence of the different properties of
intensity fluctuations for diffuse and coherent light, see Eq.~(\ref{fluc1}),
which determine the nonlinear atomic response.

In total, we obtain for the background component:
\begin{multline}
L_{\rm el}^{(2,{\rm prop})}  = \frac{\mathcal{N}}{\Sigma\ell}\int_V d{\bf
r_1}d{\bf r_2}I_\omega({\bf r_1}) I_\omega({\bf r_2})
\left|S_\omega G_\omega^{(\rm nl,a)}({\bf r_1},{\bf r_2})\right|^2\\
+\int_V \frac{d{\bf r_1}}{\Sigma\ell} I_\omega({\bf
r_1})\left(\left|G_\omega^{(\rm nl,b)}({\bf r_1})\right|^2
+\left|G_\omega^{(\rm nl,c)}({\bf r_1})\right|^2\right).
\label{nonlinlel}
\end{multline}
In the case of a slab of length $L$, Eq.~(\ref{nonlinlel}) can be simplified to:
\begin{multline}
L_{\rm el}^{(2,{\rm prop})}  =
s\int_0^L \frac{dz}{\ell} I_\omega(z)\left(2I_\omega(z)^2-2I_\omega^2(L)\right.\\
\bigl.+e^{-2z/\ell}-e^{-z/\ell}\bigr).
\label{kerrslab}
\end{multline}

\begin{figure}
\centerline{\includegraphics[width=8.5cm]{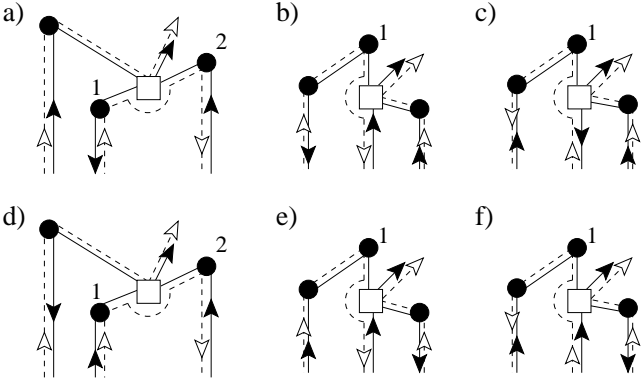}}
\caption{Interference contributions from nonlinear
propagation. The diagrams (a-f) are obtained from the ladder diagrams,
see Fig.~\ref{kerr}, by reversing the paths of the respective photons.
Just as in the case of nonlinear scattering, there are twice as many
diagrams contributing to the interference cone than to the background.
\label{kerrc}}
\end{figure}
Concerning the interference component, we find the same phenomenon which
we have already observed in the case of nonlinear scattering: if we exchange
outgoing and incoming propagators, we find \textit{twice} as many \lq
crossed\rq\ as \lq ladder\rq\ diagrams, see Fig.~\ref{kerrc}.
In particular, the diagrams (d,e,f), which could be seen as a modification
of the linear refractive index by the local
crossed intensity
- thus affecting the (ladder) average propagation -
are not considered in previously published papers, concerning either
classical linear scatterers in a non-linear
medium~\cite{cbsnl1,cbsnl2} or nonlinear scatterers in the
vacuum~\cite{wellens2,erratum}. Even if, at first sight, these diagrams look
unusual, our numerical calculations (see
section~\ref{sclass}) suggest that they play an important role, at least
in our situation where nonlinear scattering and nonlinear propagation
originate from the same microscopic process.

Due to the reciprocity symmetry (remember that nonlinear
propagation contributes to the {\em elastic} component, {\em i.e.}
no decoherence due to change of frequency ), each of the diagrams
in Fig.~\ref{kerrc} gives the same contribution as the
corresponding ladder diagram in Fig.~\ref{kerr}. Hence, to first
approximation, the interference contribution from nonlinear
propagation equals twice the background, Eq.~(\ref{nonlinlel}).
Some care must be taken, however, if photons arriving at (or
departing from) the nonlinear event (or position 1) originate from
the coherent mode. In such cases, it may happen that some of the
diagrams depicted in Figs.~\ref{kerr} and \ref{kerrc} coincide,
and we should not count them twice. [This is analogous to the
distinction between the cases (a,b,c) in Fig.~\ref{fig1}, or to
the suppression of single scattering in the linear case.]

Taking this into account [for details, we refer to the discussion after
Eq.~(\ref{polellcrossed}) in appendix \ref{appa}], we find:
\begin{multline}
C_{\rm el}^{(2,{\rm prop})}  = 2 L_{\rm el}^{(2,{\rm prop})}-
3 \int_V \frac{d{\bf r_1}}{\Sigma\ell} \left[e^{-z_1/\ell}
\left|G_\omega^{(\rm nl,c)}({\bf r_1})\right|^2\right.\\
\left.+I({\bf r_1})e^{-z_1/\ell}s\left(1-e^{-z_1/\ell}\right)\right].
\label{nonlinlelc}
\end{multline}
In the case of a slab, we obtain:
\begin{multline}
C_{\rm el}^{\rm (2,prop)}  =  2 L_{\rm el}^{(2,{\rm prop})}-3s\int_0^L
\frac{dz}{\ell} I(z)\left[e^{-z/\ell}-e^{-2b}\right]\\
+s\left(\frac{1}{2}-\frac{3}{2}e^{-2b}+e^{-3b}\right),
\end{multline}
where $b=L/\ell$ denotes the (linear) optical thickness of the slab.

Thereby, we have completed the perturbative calculation of the backscattering signal
for the scalar
case. The total signal is obtained as the sum of the various components discussed
above:
\begin{eqnarray}
L & = & L_{\rm el}^{(1)}+L_{\rm el}^{(2,{\rm scatt})}+L_{\rm el}^{(2,{\rm
prop})}+L_{\rm in}^{(2)},\label{laddertot}\\
C & = & C_{\rm el}^{(1)}+C_{\rm el}^{(2,{\rm scatt})}+C_{\rm el}^{(2,{\rm
prop})}+C_{\rm in}^{(2)}\label{crossedtot}.
\end{eqnarray}
Before we present the
numerical results in Sec.~\ref{results}, we will
generalize the above results to the
vectorial case. This is important since
polarization does not only lead to slight modifications for
low scattering orders, as in the linear case.
Apart from that, we will see that
it also induces decoherence between reversed paths,
thereby reducing the nonlinear interference components.

\subsection{Incorporation of polarization : vectorial case}
\label{pol}

First, including the polarization modifies the scalar expressions,
Eqs.~(\ref{scatt},\ref{ell}), for the
linear mean free path and the
atom-photon scattering amplitude by a factor $2/3$:
\begin{eqnarray}
\hat{\ell}&=&\left(1+\frac{4\delta^2}{\Gamma^2}\right)\frac{k^2}{6\pi{\mathcal
N}},\label{pathpol}\\
\tilde{S}_\omega & = & \frac{-6\pi i}{k(1-2i\delta/\Gamma)}\label{scattpol}
\end{eqnarray}
The Green's function, Eq.~(\ref{green}), remains unchanged, except
for the fact that the modified expression for the mean free path,
Eq.~(\ref{pathpol}), must be inserted in the refractive index.
However, the angular anisotropic character of the
atom-photon scattering is not yet contained in
Eq.~(\ref{scattpol}). This is treated by projection of the
polarization vector as follows: if the photon, with incoming
polarization $\epsilon_1$, is scattered at ${\bf r_1}$, and the next
scattering event takes place at ${\bf r_2}$, the new incoming
polarization reads:
\begin{equation}
\epsilon_2=\Delta_{{\bf r_1},{\bf r_2}}\epsilon_1,\label{proj}
\end{equation}
where $\Delta_{{\bf r_1},{\bf r_2}}$ denotes the projection onto
the plane perpendicular to ${\bf r_1}-{\bf
r_2}$. Finally, the detection signal after $n$ scattering events
is obtained as $\epsilon_D^*\cdot\epsilon_n$, with the detector
polarization $\epsilon_D$.

Thus, the linear background (\lq ladder\rq) contribution reads,
cf. Eqs.~(\ref{scattlin},\ref{iel1}):
\begin{multline}
\hat{L}^{(1,\rm el)} =  \sum_{n=1}^\infty\int\frac{d{\bf
    r_1}}{A\hat{\ell}}{\mathcal N}^{n-1}\int_V d{\bf r}_2
\dots d{\bf r}_n~e^{-z_1/\hat{\ell}}\\
\times \left(\prod_{i=1}^{n-1}|\hat{S}_\omega\hat{G}_\omega({\bf
    r_i},{\bf r_{i+1}})|^2\right) e^{-z_n/\hat{\ell}}\\
\times\frac{3}{2} \left|\epsilon_D^*\Delta_{{\bf r_{n-1}},{\bf r_n}}\dots
\Delta_{{\bf r_1},{\bf r_2}}\epsilon_L\right|^2,\label{el1pol}
\end{multline}
where $\epsilon_L$ denotes the initial laser polarization. By
choosing a given circular polarization, for example
$\epsilon_L=(1,i,0)/\sqrt{2}$, and by detecting the signal in the
helicity-preserving $h \parallel h$ polarization channel
($\epsilon_D=\epsilon_L^*$), then the single scattering
contribution in Eq.~(\ref{el1pol}) ($n=1$ term) is filtered out.
We thus recover the enhancement factor $2$, meaning $C_{\rm
el}^{(1)}=L_{\rm el}^{(1)}$. Apart from that, however,
polarization does not play a very important role: the distribution
of higher scattering orders $n>1$ is only slightly modified, and
the reciprocity symmetry remains valid, provided that
$\epsilon_D=\epsilon_L^*$.

\begin{figure}
\centerline{\includegraphics[width=7cm]{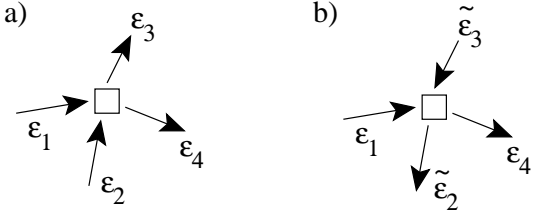}}
\caption{Polarization vectors associated to the two-photon scattering matrix
for two reversed scattering amplitudes (a) and (b). Note that the corresponding
reversed
scattering amplitudes, Eqs.~(\ref{pol1},\ref{pol2}) are different - even
in the helicity preserving polarization channel, {\em i.e.} if
$\tilde{\epsilon}_{2,3}=\epsilon_{2,3}^*$. This leads to a reduction of the CBS
interference
cone by a factor 3/4, on average.\label{polsc}
}
\end{figure}
The situation changes in the nonlinear regime of two-photon scattering.
With the initial and final polarizations $\epsilon_{1,2}$ and
$\epsilon_{3,4}$, respectively, see Fig.~\ref{polsc}(a),
the polarization dependent term of the
two-photon scattering matrix reads:
\begin{equation}
S_p=\frac{1}{2}\left[(\epsilon_1\cdot\epsilon_4^*)(\epsilon_2\cdot\epsilon_3^*)+(\epsilon_1\cdot\epsilon_3^*)(\epsilon_2\cdot\epsilon_4^*)\right].\label{pol1}
\end{equation}
The prefactor $1/2$ is chosen such that $S_p$ represents correctly
the polarized scattering amplitude in units of the corresponding
scalar one. From Eq.~(\ref{pol1}), the photon exchange symmetry
becomes evident: the outgoing photon 3, e.g., can equally well be
associated with the incoming photon 1 or 2. If we trace over the
undetected photon, which we may label as photon 4, for example, we
obtain for the ladder component:
\begin{multline}
  \Pi^{(L)}(\epsilon_1,\epsilon_2;\epsilon_3) = \sum_{\epsilon_4}
|S_p|^2=\frac{1}{4}\Bigl[
|\epsilon_2\cdot\epsilon_3^*|^2+|\epsilon_1\cdot\epsilon_3^*|^2\Bigr.\\
\left.+2 {\rm
  Re}\Bigl\{(\epsilon_1\cdot\epsilon_2^*)(\epsilon_2\cdot\epsilon_3^*)(\epsilon_3\cdot\epsilon_1^*)\Bigr\}\right].\label{poll}
\end{multline}
If we assume a random uniform distribution for the polarization
vectors, we obtain $\langle \Pi^{(L)}\rangle=2/9$, which is
smaller than the linear counterpart
$\langle|\epsilon_n\cdot\epsilon_D^*|^2\rangle=1/3$. Hence, in the
vectorial case, the relative weight of the nonlinear contribution
is approximately one third smaller than in the scalar case - at
least far inside the medium, where the polarization is
sufficiently randomized.

Concerning the interference (\lq crossed\rq) contribution, we
exchange the direction of the outgoing detected photon 3 and of
one of the incoming photons, for example photon 2. Note that we
obtain in general different polarizations $\tilde{\epsilon}_{2,3}$
for the reversed counterparts of $\epsilon_{2,3}$, see
Fig.~\ref{polsc}(b). Indeed, the reversed photons have the same
polarizations, $\tilde{\epsilon}_{2,3}=\epsilon_{2,3}$, only if
the laser and detector polarizations are identical
($\epsilon_D=\epsilon_L$). Consequently, the scattering amplitude
for the complex conjugate photon pair reads:
\begin{equation}
\tilde{S}_p=\frac{1}{2}\left[(\epsilon_1\cdot\epsilon_4^*)
(\tilde{\epsilon}_3\cdot\tilde{\epsilon}_2^*)+
(\epsilon_1\cdot\tilde{\epsilon}_2^*)(\tilde{\epsilon}_3\cdot\epsilon_4^*)
\right].\label{pol2}
\end{equation}
Note that even in the helicity-preserving polarization channel,
\emph{i.e.} $\tilde{\epsilon}_{2,3}=\epsilon_{2,3}^*$, the
reversed scattering amplitudes, Eqs. (\ref{pol1},\ref{pol2}), are
in general not equal. Only the first term, where photon 2 is
associated with photon 3, remains unchanged if those two photons
are reversed. As a consequence, the polarization induces a loss of
coherence, \emph{i.e.} a reduction of the crossed term as compared
to the scalar case. The sum over the polarization of photon 4
yields:
\begin{multline}
\Pi^{(C)}(\epsilon_1,\epsilon_2,\tilde{\epsilon}_3,\epsilon_3,\tilde{\epsilon}_2) =
\sum_{\epsilon_4} S_p\tilde{S}_p^*=\\
=\frac{1}{4}\left[(\epsilon_2\cdot\epsilon_3^*)(\tilde{\epsilon}_2\cdot\tilde{\epsilon}_3^*)
+(\epsilon_2\cdot\epsilon_3^*)(\epsilon_1\cdot\tilde{\epsilon}_3^*)
(\tilde{\epsilon_2}\cdot\epsilon_1^*)\right.\\
\left.+
(\epsilon_1\cdot\epsilon_3^*)(\epsilon_2\cdot\tilde{\epsilon}_3^*)
(\tilde{\epsilon_2}\cdot\epsilon_1^*)
+(\epsilon_1\cdot\epsilon_3^*)(\epsilon_2\cdot\epsilon_1^*)
(\tilde{\epsilon_2}\cdot\tilde{\epsilon_3}^*)\right].\label{polc}
\end{multline}
If we assume $\tilde{\epsilon}_{2,3}=\epsilon^*_{2,3}$,
\emph{i.e.} the $h\parallel h$ channel, we obtain
$\langle\Pi^{(C)}\rangle=3/18$ on average. Hence, in this channel,
the polarization-induced loss of contrast is approximately
$\langle\Pi^{(C)}\rangle/\langle\Pi^{(L)}\rangle=3/4$.

\begin{figure}
\centerline{\includegraphics[width=7cm]{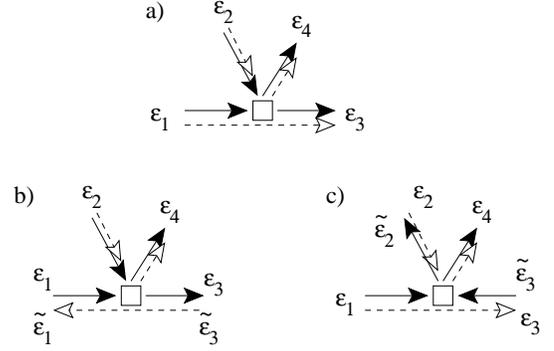}}
\caption{Polarization dependence of nonlinear propagation
for (a) ladder and (b,c) crossed diagrams.
\label{polpr}}
\end{figure}
Finally, to obtain the polarization dependence of nonlinear propagation,
we label the photons as shown in Fig.~\ref{polpr}.
Let us first examine the ladder term, Fig.~\ref{polpr}(a).
The solid lines are described by the
two-photon amplitude, Eq.~(\ref{pol1}), whereas
the dashed lines give the complex conjugate of
$(\epsilon_2\cdot\epsilon_4^*)(\epsilon_1\cdot\epsilon_3^*)$.
After integration over photon 4, the result is
\begin{equation}
\begin{aligned}
\Pi^{(L,\text{prop})}(\epsilon_1,\epsilon_2,\epsilon_3)  = &
\frac{1}{2}\left[(\epsilon_1\cdot\epsilon_2^*)(\epsilon_2\cdot\epsilon_3^*)
(\epsilon_3\cdot\epsilon_1^*)\right.\\
& +\left.(\epsilon_1\cdot\epsilon_3^*)(\epsilon_2\cdot\epsilon_2^*)
(\epsilon_3\cdot\epsilon_1^*)\right].\label{polpath}
\end{aligned}
\end{equation}
Concerning the crossed diagrams, we distinguish between the two cases shown
in Fig.~\ref{polpr}(b) and (c). [In Fig.~\ref{kerrc},
this corresponds to (a,b,c), on the one hand,
and (d,e,f), on the other hand.] As for the case (b), nothing changes since
the reversed photon does not participate in the nonlinear event.
In case (c), we obtain:
\begin{multline}
\Pi^{(C,\text{prop})}(\epsilon_1,\epsilon_2,\epsilon_3,\tilde{\epsilon}_2,\tilde{\epsilon}_3)=\frac{1}{2}\left[(\epsilon_1\cdot\epsilon_2^*)(\tilde{\epsilon_3}\cdot\tilde{\epsilon_2}^*)
(\epsilon_3\cdot\epsilon_1^*)\right.\\
+\left.(\epsilon_1\cdot\tilde{\epsilon_2}^*)(\tilde{\epsilon_3}\cdot\epsilon_2^*)
(\epsilon_3\cdot\epsilon_1^*)\right].\label{polpathc}
\end{multline}
When determining the average values of the nonlinear propagation terms,
it must be taken into account
that $\epsilon_1$ and $\epsilon_3$ are not independent from each other, since
they propagate in the same (or opposite) direction. Thus, we find
$\langle \Pi^{(L,\text{prop})}\rangle =1/3$ and $\langle
\Pi^{(C,\text{prop})}\rangle =1/6$.
Hence, the loss of contrast equals $1/2$ in case (c), whereas reciprocity
remains conserved ({\em i.e.} no loss of contrast) in case (b). Averaging over (b)
and (c),
this yields the same contrast $3/4$ as for nonlinear scattering.

What remains to be done to obtain the vectorial backscattering signal
is to incorporate the above expressions into
the corresponding scalar equations. The resulting equations can be found in
appendix~\ref{appa},
together with a description of the Monte-Carlo method which we use for their
numerical solution.

\subsection{Classical model}
\label{sclass}

We want to stress that our perturbative
theory of nonlinear coherent backscattering
is not only valid for an atomic medium, but can be adapted
to other kinds of nonlinear scatterers. In particular, the effect of
interference between
three amplitudes is always present in the perturbative regime of a
small $\chi^{(3)}$ nonlinearity.
Specifically, we have also examined the following model:
a collection of classical isotropic
scatterers, situated at positions ${\bf r_i}$, $i=1,\dots,N$.
In analogy to the atomic model,
we assume that the field scattered elastically by an individual
scatterer at position
${\bf r_i}$ is proportional
to $E_i/(1+s|E_i|^2)$, where $E_i$ is the local field at ${\bf r_i}$,
and $s$ measures the strength of the nonlinearity.
Writing $E_i$ as a sum of the incident field, and the field radiated by
all other scatterers, we obtain the following set of nonlinear equations:
\begin{equation}
E_i=e^{i{\bf k_L}\cdot {\bf r_i}}+i \sum_{j\neq i}\frac{e^{i k
    r_{ij}}}{kr_{ij}}
\frac{E_j}{1+s|E_j|^2}.\label{class}
\end{equation}

Employing diagrammatic theory similar to the one outlined above, we have
checked that, in the ensemble average over the positions ${\bf r_i}$,
this model indeed reproduces the
{\em elastic} components of the backscattering signal of the atomic model.
We have checked that the results obtained from direct
numerical solutions of the field equations (\ref{class}) -
averaged over a sufficiently large sample of single realizations -
agree with our theoretical predictions, in the perturbative regime of small
nonlinearity $s$. In particular, the diagrams (d,e,f) of
Fig.~\ref{kerrc}, describing the interference contributions from the
nonlinear propagation, are essential to give the correct results.
A more detailed analysis will be presented elsewhere.

Furthermore, it remains to be clarified whether the diagrams
(d,e,f) are also relevant for the description of propagation in
{\em homogeneous} nonlinear media, into which linear scatterers
are embedded at random positions. First studies of the resulting
CBS cone have been presented in \cite{cbsnl1,cbsnl2}, without
taking into account interference between three amplitudes,
however. Experimentally, this question can be resolved by
measuring the value of the backscattering enhancement factor
$\eta$: whereas $\eta$ is basically unaffected by the nonlinearity
according to \cite{cbsnl1,cbsnl2} ({\em i.e.} $\eta=2$
apart from single scattering), our equations
(\ref{nonlinlel},\ref{nonlinlelc}), with $s$ proportional to the
incoming intensity and to the $\chi^{(3)}$ coefficient of the
nonlinear Kerr medium, predict a significant change of $\eta$ when
varying the incoming intensity.

\section{Results}
\label{results}
We return to the atomic model, concentrating on the case of a slab geometry
in the following.
 Using the equations derived in Sec.~
\ref{slin}-\ref{pol},
we are able to calculate the backscattered intensity
up to first order in the saturation parameter $s$.
In this section,
we will examine its
dependence on the optical thickness $b$ and detuning $\delta$, for the scalar and
vectorial case.
The main quantity of interest
is the backscattering enhancement factor $\eta$. It is defined as the ratio between the
total detection signal in exact backscattering direction divided by the background
component.
If we perform an expansion up to first order in $s$, we obtain
\begin{equation}
\eta  = \frac{L+C}{L}
\simeq \eta^{(1)}+(\eta^{(1)}-1)(\gamma_C-\gamma_L)s.\label{alpha}
\end{equation}
Here, $\eta^{(1)}=1+C_{\rm el}^{(1)}/L_{\rm el}^{(1)}$ is the
enhancement factor in the linear case (\emph{i.e.} the limit of
vanishing saturation). If single scattering is excluded (e.g. in
the $h\parallel h$ channel), we have $\eta^{(1)}=2$. Increasing
saturation changes the enhancement factor, and the present
approach allows us to calculate the slope $d\eta/ds$ of this
change at $s=0$. It is given by the difference between the
nonlinear crossed and ladder contribution, normalized as follows:
\begin{eqnarray}
\gamma_L & = & \frac{L-L^{(1)}}{s L^{(1)}}\label{gl},\\
\gamma_C & = & \frac{C-C^{(1)}}{s C^{(1)}}\label{gc}.
\end{eqnarray}

Obviously, an important question is the domain of validity of the
linear expansion, Eq.~(\ref{alpha}). Strictly speaking, this
question can only be answered if we know higher orders of $s$.
However, a rough quantitative estimation can be given as follows:
if $p_1$ (resp. $p_{2+}$) denotes the probability for a
backscattered photon to undergo one (resp. more than one)
nonlinear scattering event, the perturbative condition reads
$p_{2+}\ll p_1$. If we assume that all scattering events
have the same probability (proportional to $s$) to be nonlinear
(thereby neglecting the inhomogeneity of the local intensity), we
obtain $p_1\simeq \langle N\rangle s$, and $p_{2+}\simeq \langle
N^2\rangle s^2$, where $N$ denotes the total number of scattering
events, and $\langle\dots\rangle$ the statistical average over all
backscattering paths. Evidently, $N$ and $N^2$ are expected to
increase when increasing the optical thickness $b$. For a slab
geometry, we have found numerically that $\langle N\rangle\propto
b$ and $\langle N^2\rangle\propto b^3$ (in the limit of large
$b$), concluding that the perturbative treatment is valid if
$sb^2\ll 1$. Let us note that a similar condition also ensures the
stability of speckle fluctuations in a nonlinear medium
\cite{skipetrov}.

\begin{figure}
\centerline{\includegraphics[height=8cm,angle=270]{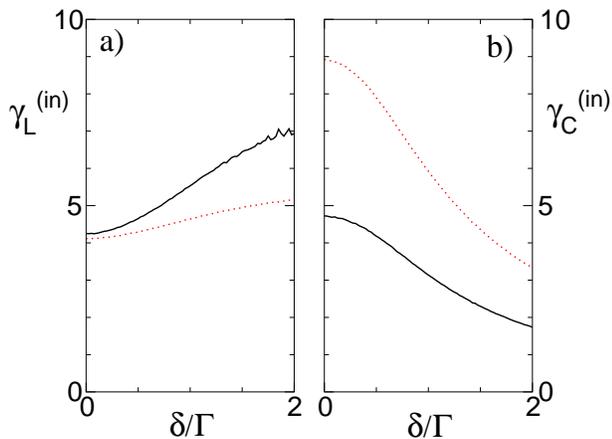}}
\caption{(color online). Normalized inelastic ladder and crossed contributions
$\gamma_{L,C}^{(\rm in)}$, cf. Eqs.~(\ref{gl},\ref{gc}), for
optical thickness $b=0.5$, as a function of the laser detuning
$\delta$. Solid lines: polarized case ($h||h$ channel). Dotted
lines: scalar case. For comparison, the corresponding elastic
contributions (independent of $\delta$) are $\gamma_L^{(\rm
el)}=-7.04$ ($h\parallel h$), $-6.53$ (scalar), and
$\gamma_C^{(\rm el)}=-9.56$ ($h\parallel h$), $-18.8$ (scalar).
\label{num1}}
\end{figure}
In Fig.~\ref{num1}, we show the inelastic ladder and crossed
contributions $\gamma_L^{(\rm in)}$ and $\gamma_C^{(\rm in)}$ for a
slab of optical thickness $b=0.5$ as a function of the detuning,
$\delta=\omega-\omega_{\rm at}$,
for the polarized ($h\parallel h$) and scalar case.
Since the optical thickness is kept constant, the elastic
quantities are independent of the detuning, and only the inelastic
components are affected by $\delta$, via the shape of the
power spectrum $P(\omega')$ of the inelastically scattered light,
see Eq.~(\ref{specin}).  The latter exhibits two peaks of width
$\Gamma$, one of which is centered around the atomic resonance.
The increase of the ladder term as a function of $\delta$, which
is observed in Fig.~\ref{num1}(a) is due to initially detuned
photons, {\em i.e.} $\omega=\omega_{\rm at}+\delta$, which
are set to resonance ($\omega'\simeq \omega_{\rm at}$) by the
nonlinear scattering process. For these photons, the scattering
cross section increases, which increases the contribution to the
backscattering signal in the sum over all scattering orders -
especially in the $h\parallel h$ case where single scattering is
filtered out. The same effect also applies for the crossed term,
Fig.~\ref{num1}(b), but here the dephasing between the reversed
paths due to the frequency change - which is more effective for
higher values of the detuning - is dominant, leading in total to a
decrease of $\gamma_C^{(\rm in)}$ as a function of $\delta$.
The small ripples in Fig.~\ref{num1}(a), for the polarized case (solid line) 
at large $\delta$, are due to numerical noise in the Monte-Carlo integration.

\begin{figure}
\centerline{\includegraphics[height=8cm,angle=270]{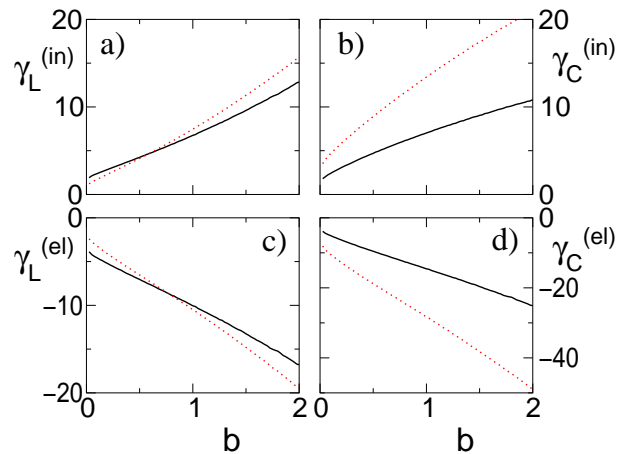}}
\caption{(color online). Normalized inelastic and elastic ladder and crossed
$\gamma_{L,C}^{(\rm el,in)}$, cf. Eqs.~(\ref{gl},\ref{gc}),
for vanishing detuning $\delta=0$, as a function of the
optical thickness $b=0.5$. Solid lines: polarized case ($h||h$ channel).
Dotted lines: scalar case.\label{d0}}
\end{figure}
Fig.~\ref{d0} shows the elastic and inelastic ladder and crossed
contributions, as a function of the optical thickness, at detuning
$\delta=0$. The main purpose of this figure is to show the
increase of the nonlinear contributions as a function of $b$,
which is important to understand the domain of validity of the
present approach. The origin of this increase is simple to
understand: for larger values of the optical thickness, the
average number of scattering events increases, and so does also
the probability that at least one of them is a nonlinear one.
Thus, for an optically thick medium, even a very small initial
saturation may lead to a large inelastic component of the
backscattered light. Note, however, that the elastic and inelastic
ladder contributions, Fig.~\ref{d0}(a,c), tend to cancel each
other, such that their sum depends less strongly on $b$.
Physically, this fact is related to energy conservation.
The latter ensures that the {\em total} nonlinear scattered
intensity - integrated over all final directions - vanishes even
exactly, since the total outgoing intensity must equal
the incident intensity (meaning a purely linear
relationship between outgoing and incident intensity).

Furthermore, we note that
both the elastic and inelastic
ladder components increase significantly
slower in the polarized than in the scalar case (solid vs. dashed line). This is due
to the
fact that, as discussed in Sec.~\ref{pol}, polarization effects diminish
the weight of nonlinear scattering by approximately 2/3.
Concerning the crossed components, Fig.~\ref{d0}(b,d), the difference is even stronger,
due to the additional polarization-induced loss of contrast by a factor $3/4$, in
average.
Please note that the vertical scale for the elastic crossed case, Fig.~\ref{d0}(d),
is two times larger than in the other three cases:
this reflects the effect of interference between three amplitudes, which renders the
crossed
component up to two times larger than the ladder. Concerning the inelastic
component, Fig.~\ref{d0}(b), this effect is diminished by decoherence due to the
frequency change
at inelastic scattering. Here, crossed and ladder component are of similar magnitude.

\begin{figure}
\centerline{\includegraphics[height=8cm,angle=270]{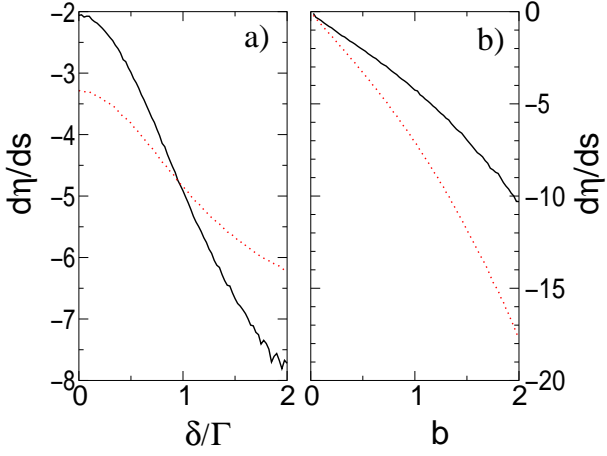}}
\caption{(color online). Slope of backscattering enhancement factor, for the
parameters of Fig.~\ref{num1} ($b=0.5$, left half) and Fig.~\ref{d0}
($\delta=0$, right half). Solid lines: polarized case ($h||h$ channel).
Dotted lines: scalar case.\label{num2}}
\end{figure}
In Fig.~\ref{num2}, we show the slope of the backscattering
enhancement factor, which follows via Eq.~(\ref{alpha}) from
the data shown in Figs.~\ref{num1} and \ref{d0}.
Fig.~\ref{num2}(b) again points out the importance of even small
saturation in the case of an optically thick medium. For example,
in the scalar case at $b=2$, increasing the saturation
from $s=0$ to $s=0.01$ decreases the
enhancement factor from $1.73$ ($<2$ due to single scattering) to $1.55$.
For very large $b$, we find a linear decrease of the
slope. At the same time, however, the
allowed domain of $s\ll  1/b^2$ shrinks to zero {\em quadratically}.
This allows the enhancement factor to remain a continuous function of $s$, even in
the limit
$b\to\infty$, where its slope at $s=0$ diverges.
In order to make more precise statements about the behavior in the
limit $b\to\infty$, however, it is necessary to generalize our theory
to the case of more than one nonlinear scattering event.

\begin{figure}
\centerline{\includegraphics[height=8cm,angle=270]{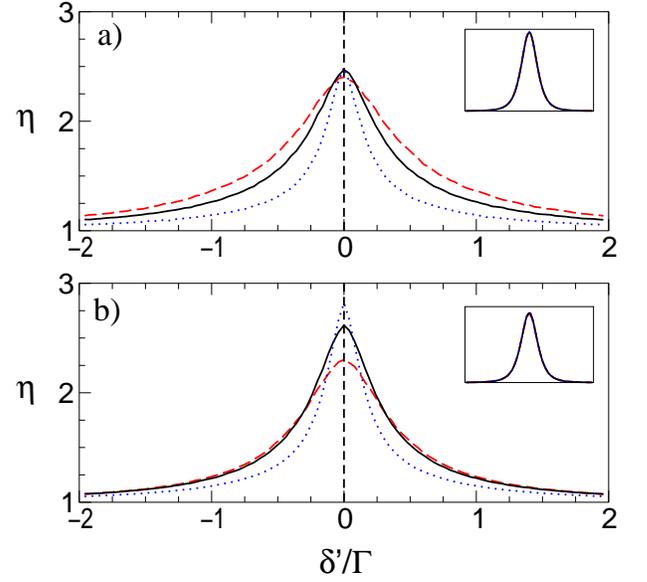}}
\caption{(color online). Spectral dependence of the enhancement factor, for
detuning $\delta=0$ and optical thickness $b=0.5$ (dashed line),
$1$ (solid), and $2$ (dotted), in the $h\parallel h$ channel (a)
and the scalar case (b). The vertical dashed line
displays the position of the elastic $\delta$ peak, which must be
filtered out in order to observe an enhancement factor larger than
two. The inset shows the power spectrum of the backscattered light
(background component), which is almost identical with the
single-atom spectrum. \label{spek3}}
\end{figure}
\begin{figure}
\centerline{\includegraphics[height=8cm,angle=270]{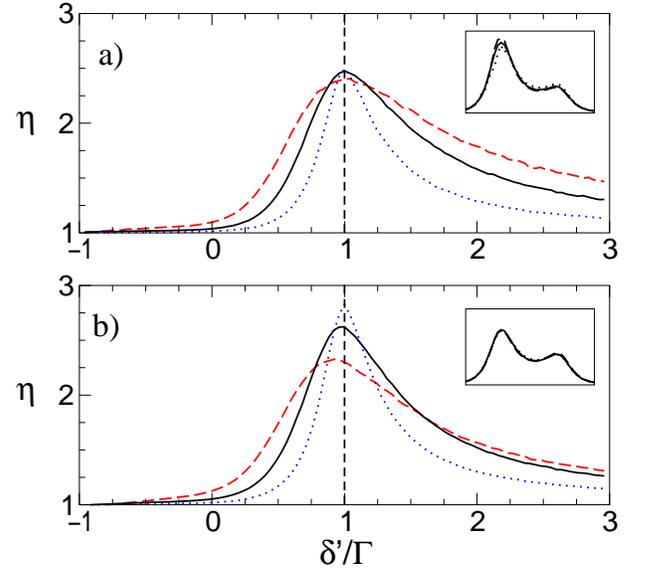}}
\caption{(color online). Spectral dependence of the enhancement factor, for
detuning $\delta=\Gamma$ and optical thickness $b=0.5$ (dashed
line), $1$ (solid), and $2$ (dotted), in the $h\parallel h$
channel (a) and the scalar case (b). The vertical dashed
line displays the position of the elastic $\delta$ peak, which
must be filtered out in order to observe an enhancement factor
larger than two. The inset shows the power spectrum of the
backscattered light (background component), revealing the
amplification of the on-resonance peak with respect to the
symmetric single-atom spectrum. \label{spek4}}
\end{figure}

On the left hand side, Fig.~\ref{num2}(a) depicts the dependence of the
enhancement factor on detuning, for $b=0.5$. As already discussed above, the
decrease of $\eta$ with increasing $\delta$ originates
from the form of the inelastic power spectrum, which results in a stronger
dephasing between reversed paths for larger detuning.
Thus,
the modification of the enhancement factor with the detuning,
keeping fixed the linear optical thickness, is a signature of the
nonlinear atomic response and has been experimentally observed in
ref.~\cite{thierry}.
Let us stress, however,
that in the cases shown in Figs.~\ref{num1} and \ref{num2}(a) and small
detuning, the inelastic component gives a {\em positive}
contribution to the backscattering enhancement factor. Hence, the
observed negative slope of $\eta$ originates from the
{\em elastic} component, where the nonlinear crossed term is up to two times larger
than the ladder, but with negative sign, see Fig.~\ref{d0}(c,d).

In order to observe an enhancement factor larger than two - and
thereby demonstrate clearly the effect of interference between
three amplitudes - it is therefore necessary to filter out the
elastic component. In principle, this can be achieved by means of
a spectral filter, \emph{i.e.} by detecting only photons with a
certain frequency $\omega'$, different from the laser frequency
$\omega$. Thereby, it is possible to measure the spectral
dependence of the backscattering enhancement factor, see
Fig.~\ref{spek3}. Here, the upper (a) and lower (b) half depict
the polarized ($h\parallel h$) and scalar case, respectively, for
vanishing laser detuning, $\delta=0$. Evidently, the largest
values of the enhancement factor are obtained if the final
frequency approaches the initial one, since then the dephasing due
to different frequencies vanishes. In the scalar case, the value
of the enhancement factor in the limit $\delta'\to 0$ is
completely determined by the relative weights between the one-,
two- and three-amplitudes cases shown in Fig.~\ref{fig1}, cf.
Eq.~(\ref{laddcross}). As evident from the dashed line in
Fig.~\ref{spek3}(b), already at the rather moderate value $b=0.5$
of the optical thickness, the three-amplitudes case is
sufficiently strong in order to increase the maximum enhancement
factor above the linear barrier $\eta=2$. With increasing optical
thickness (and, if necessary, decreasing saturation parameter, in
order to stay in the domain of validity of the perturbative
approach, see above), the number of linear scattering events
increases, which implies that the three-amplitudes case
increasingly dominates, see Fig.~\ref{fig1}. In this limit, the
enhancement factor approaches the maximum value three. At the same
time, however, a larger number of scattering events also leads to
stronger dephasing due to different frequencies,
$\omega'\neq\omega$. This results in a narrower shape of $\eta$ as
a function of $\omega'$ for larger optical thickness.
Nevertheless, as evident from Fig.~\ref{spek3}(b), the enhancement
factor remains larger than two in a significant range of
frequencies $\omega'$. The same is true for the polarized case,
Fig.~\ref{spek3}(a). However, here the enhancement factor cannot
exceed the value $2.5$, due to the polarization-induced loss of
contrast. At the same time, the optical thickness has less
influence on the maximum enhancement factor at $\delta'=0$, since
single scattering, Fig.~\ref{fig1}(a) - and partly also the
two-amplitudes case, Fig.~\ref{fig1}(b) - are filtered out, so
that interference of three amplitudes already prevails at rather
small values of the optical thickness.

In Fig.~\ref{spek4}, the influence of an initial detuning (here:
$\delta=\Gamma$)
is displayed. Basically, the above conclusions remain almost equally
valid for the
detuned case. A small difference is seen in the scalar case,
Fig.~\ref{spek4}(b), where the
maximum of $\eta(\delta')$ is found slightly below $\delta$. This is
due to the fact
that the weight of single scattering increases with increasing
$\delta'$. Furthermore, the
inset reveals that the power spectrum of the backscattered light
differs from the
single-atom spectrum, Eq.~(\ref{specin}), where the two peaks at
$\delta'=0$ and
$\delta'=2 \delta$ are equally strong. In the multiple scattering
case, the on-resonance
peak at $\delta'=0$ is amplified, since the scattering cross-section is larger for
photons on resonance. As already mentioned above, see the discussion
of Fig.~\ref{num1},
this increases the total contribution to the detection signal (in the
sum over all scattering paths) -
especially in the polarized case, where single scattering is filtered out.

\section{Conclusion}
\label{concl}

In summary, we have presented a detailed diagrammatic calculation
of coherent backscattering of light
from a dilute medium composed of weakly saturated two-level atoms.
Our theory applies in the perturbative two-photon scattering regime
($s\ll1$ and $sb^2\ll 1$), where at most one nonlinear scattering event occurs.
The value of the backscattering enhancement
factor is determined by the following three effects: firstly, due to the
nonlinearity of the atom-photon interaction,
there may be either two or three different amplitudes which interfere in
backscattering direction. This implies
a maximal enhancement factor between two and three
for the nonlinear component, where the value three is approached
for large optical thickness.
However, since the contribution from
nonlinear scattering has a negative sign, the total enhancement
factor (linear plus nonlinear elastic and inelastic components) is
{\em reduced}
by the effect of three-amplitudes interference.
Only if the elastic
component is filtered out, a value larger than two can be observed.

Secondly, a loss of coherence is implied by the change of
frequency due to inelastic scattering - like in the case of two
atoms \cite{wellens}. The random frequency change leads to
different scattering phases - and hence on average decoherence -
between reversed paths. Finally, a further loss of contrast is
induced by nonlinear polarization effects - even in the
$h\parallel h$ channel, which exhibits ideal contrast in the
linear case. Nevertheless, the enhancement factor remains larger
than two in certain frequency windows of the inelastic
backscattering signal. Thus, it is experimentally possible to
clearly identify the effect of interference between three
amplitudes - provided a sufficiently narrow spectral filter is at
hand.

A natural way to extend this work is to give up the
perturbative assumption, and admit more than one nonlinear
scattering event. This is necessary in order to describe media
with large optical thickness, even at small saturation. Since the
number of interfering amplitudes increases if more than two
photons are connected by nonlinear scattering events, we expect
the occurrence of even larger enhancement factors in the
nonperturbative regime - especially in the case of scatterers with
positive nonlinearity, {\em i.e.} for scatterers whose cross
section {\em increases} with increasing intensity.

Furthermore, the relation between coherent backscattering and weak localization
in the presence of nonlinear scattering remains to be explored.
Does a large enhancement of coherent backscattering also imply a strong reduction
of nonlinear diffusive transport? If the answer is yes - as it is the case in the
linear regime - this implies that
wave localization can be facilitated by introducing appropriate nonlinearities.

\acknowledgements
T.W. was supported by the DFG Emmy Noether program.
Laboratoire Kastler Brossel is laboratoire de l'Universit\'e Pierre et Marie
Curie et de l'Ecole Normale Superieure, UMR 8552 du CNRS.

\appendix

\section{Monte-Carlo simulation}
\label{appa}

As discussed in Sec.~\ref{pol}, the incorporation of polarization effects requires to
take into account the projection of polarization vectors in the corresponding scalar
equations.
For the inelastic ladder component, insertion of the polarization term,
Eq.~(\ref{poll}),
into the scalar expression, Eq.~(\ref{lnonlin}), yields:
\begin{multline}
\hat{L}_{\rm in}^{(2)} =  s\int\frac{d{\bf r}}{A\hat{\ell}}\int
d\omega' P(\omega')\sum_{n,m,l=0}^\infty \mathcal{N}^{n+m+l}
\times\\
\int_V d{\bf u_1}\dots d{\bf u_n}\quad
e^{-u_{1,z}/\hat{\ell}}\left(\prod_{i=1}^n
\left|\hat{S}_\omega \hat{G}_\omega({\bf u_i},{\bf u_{i+1}})\right|^2\right)\\
\int_V d{\bf v_1}\dots d{\bf v_m}\quad
e^{-v_{1,z}/\hat{\ell}}\left(\prod_{j=1}^m
\left|\hat{S}_\omega \hat{G}_\omega({\bf v_j},{\bf v_{j+1}})\right|^2\right)\\
\int_V d{\bf w_1}\dots d{\bf w_l}\quad
e^{-w_{1,z}/\hat{\ell}'}\left(\prod_{k=1}^l
\left|\hat{S}_{\omega'} \hat{G}_{\omega'}({\bf w_k},{\bf
    w_{k+1}})\right|^2\right)\\
\frac{3}{2}\Pi^{(L)}(\epsilon_u,\epsilon_v;\epsilon_w)\times
\begin{cases}
1 & \text{ if } n=m=0,\\
2 & \text{ if } n>0 \text{ or } m>0,
\end{cases}
\label{ladderpol}
\end{multline}
with ${\bf u_{n+1}}={\bf v_{m+1}}={\bf w_{l+1}}={\bf r}$. Furthermore,
the polarization vectors are given by:
\begin{equation}
\begin{aligned}
\epsilon_u & =   \Delta_{{\bf u_n},{\bf u_{n+1}}}\dots
\Delta_{{\bf u_1},{\bf u_2}}\epsilon_L,\\
\epsilon_v & =   \Delta_{{\bf v_m},{\bf v_{m+1}}}\dots
\Delta_{{\bf v_1},{\bf v_2}}\epsilon_L,\\
\epsilon_w & =   \Delta_{{\bf w_l},{\bf w_{l+1}}}\dots
\Delta_{{\bf w_1},{\bf w_2}}\epsilon_D.\\
\end{aligned}
\end{equation}
The analogous procedure for the interference component, inserting
Eq.~(\ref{polc}) into
Eq.~(\ref{nonlinc}), yields:
\begin{multline}
\hat{C}^{\rm (2,in)} = s\int \frac{d{\bf r}}{A\hat{\ell}}\int
d\omega' P(\omega')\sum_{n,m,l=0}^\infty
{\mathcal N}^{n+m+l}\\
\int_V d{\bf u_1}\dots d{\bf w_l}\quad
 e^{-u_{1,z}/\hat{\ell}}\left(\prod_{i=1}^n
\left|\hat{S}_\omega \hat{G}_\omega({\bf u_i},{\bf u_{i+1}})\right|^2
\right)\\
e^{ikv_{1,z}(n_\omega+n^*_{\omega'})}
\left(\prod_{j=1}^m
\hat{S}_\omega\hat{S}^*_{\omega'}\hat{G}_\omega({\bf v_j},{\bf v_{j+1}})
\hat{G}^*_{\omega'}({\bf v_j},{\bf v_{j+1}})
\right)\\
e^{ikw_{1,z}(n^*_\omega+n_{\omega'})}
\left(\prod_{k=1}^l
\hat{S}^*_\omega\hat{S}_{\omega'}\hat{G}^*_\omega({\bf w_k},{\bf w_{k+1}})
\hat{G}_{\omega'}({\bf w_k},{\bf w_{k+1}})
\right)\\
\frac{3}{2}
\Pi^{(C)}(\epsilon_u,\epsilon_v,\tilde{\epsilon}_w,\epsilon_w,\tilde{\epsilon}_v)\times
\begin{cases}
0, & \text{if } m=l=0,\\
2, & \text{if }n=m=0,\quad l>0\\
2, & \text{if }n=l=0,\quad m>0,\\
4, & \text{otherwise}
\end{cases}\label{crossedpol}
\end{multline}
with the polarization vectors of the \lq reversed\rq\ photons:
\begin{equation}
\begin{aligned}
\tilde{\epsilon_v} & =  \Delta_{{\bf v_m},{\bf v_{m+1}}}\dots
\Delta_{{\bf v_1},{\bf v_2}}\epsilon_D,\\
\tilde{\epsilon}_w & =   \Delta_{{\bf w_l},{\bf w_{l+1}}}\dots
\Delta_{{\bf w_1},{\bf w_2}}\epsilon_L.
\end{aligned}
\end{equation}
The elastic nonlinear scattering components follow simply
by inserting $-2\delta(\omega'-\omega)$ instead of the inelastic power spectrum
$P(\omega')$
in the above Eqs.~(\ref{ladderpol},\ref{crossedpol}).

The nonlinear propagation term is obtained by inserting
Eq.~(\ref{polpath}) into Eq.~(\ref{nonlinlel}):
\begin{multline}
\hat{L}_{\rm el}^{\rm (2,prop)} = s\sum_{n=1}^\infty{\mathcal N}^n
\int_V d{\bf u_1}\dots d{\bf u_n}\quad
 e^{-(u_{1,z}+u_{n,z})/\hat{\ell}}\\
\left(\prod_{i=1}^n
\left|\hat{S}_\omega\hat{G}_\omega({\bf u_i},{\bf u_{i+1}})\right|^2\right)
\sum_{m=1}^\infty {\mathcal N}^{m-1}\int_V d{\bf v}_1\dots d{\bf v}_{m-1}\\
\sum_{l=0}^n \int_{\bf u_l}^{\bf u_{l+1}}\frac{d{\bf v_m}}{\hat{\ell}}
\quad e^{-v_{1,z}/\hat{\ell}}\left(\prod_{i=1}^{m-1}
\left|\hat{S}_\omega\hat{G}_\omega({\bf v_i},{\bf v_{i+1}})\right|^2
\right)\\
\Pi^{(L,\text{prop})}(\epsilon_1,\epsilon_v,\epsilon_3)\times
\begin{cases}
1 & \text{if } m=l=0,\\
2 & \text{otherwise},
\end{cases}\label{ellpol}
\end{multline}
Here, the nonlinear event takes place between $\bf u_l$ and $\bf u_{l+1}$.
Correspondingly, $\int_{\bf u_l}^{\bf u_{l+1}}$ denotes the one-dimensional
integral on a straight line between these points, and
${\bf u_0}={\bf u_1}-u_{1,z} {\bf e_z}$ and ${\bf u_{n+1}}={\bf
  u_n}-u_{n,z}{\bf e_z}$
are defined as the points where the photon
enters or leaves the medium, respectively.
The three cases Fig.~\ref{kerr}(a,b,c)
correspond to $0<l<n$, $l=n$, and $l=0$, respectively.
The polarization vectors $\epsilon_1$ and $\epsilon_3$ participating in the
nonlinear event,
cf. Fig.~\ref{polpr}, are obtained as:
\begin{equation}
\begin{aligned}
\epsilon_1 & =   \Delta_{{\bf u_{l+1}},{\bf u_l}}\dots
\Delta_{{\bf u_2},{\bf u_1}}\epsilon_L,\\
\epsilon_3 & =   \Delta_{{\bf u_l},{\bf u_{l+1}}}\dots
\Delta_{{\bf u_{n-1}},{\bf u_n}}\epsilon_D.
\end{aligned}
\end{equation}
Finally, to obtain the interference component $\hat{C}^{\rm
  (2,prop)}_{\rm el}$,
the last term in Eq.~(\ref{ellpol}) must be replaced by:
\begin{multline}
\Pi^{(L,\text{prop})}(\epsilon_1,\epsilon_v,\epsilon_3)\times
\begin{cases}
2 & \text{if } n>1, (m,l)\neq(0,0),\\
1 & \text{if } n>1, m=l=0,\\
0 & \text{otherwise}
\end{cases}+\\
\Pi^{(C,\text{prop})}(\epsilon_1,\epsilon_v,\epsilon_3,
\tilde{\epsilon}_v,\tilde{\epsilon}_3)\times
\begin{cases}
4 & \text{if } l=0, n>1, m>0,\\
2 & \text{if } l=0, n=1, m>0,\\
2 & \text{if } 0<l<n,\\
0 & \text{otherwise},
\end{cases}\label{ellpolc}
\end{multline}
with
\begin{equation}
\tilde{\epsilon}_3  =   \Delta_{{\bf u_l},{\bf u_{l+1}}}\dots
\Delta_{{\bf u_{n-1}},{\bf u_n}}\epsilon_L.\label{polellcrossed}
\end{equation}
The first term, $\Pi^{(L,{\rm prop})}$,
equals the ladder component minus single scattering ($n=1$), whereas the
second one, $\Pi^{(C,{\rm prop})}$,
describes the additional crossed diagrams shown in Fig.~\ref{kerrc}(d-f).
Here, the case $0<l<n$ corresponds to Fig.~\ref{kerrc}(d), where the
nonlinearity occurs between two scattering events. The remaining diagrams,
Fig.~\ref{kerrc}(e,f), correspond to $l=0$. Here, the case $m=0$ (\lq
pump photon from the
coherent mode\rq) does not contribute, since then the diagrams
Fig.~\ref{kerrc}(e,f) are
identical to Fig.~\ref{kerrc}(b,c). Furthermore, if $n=1$ (\lq probe
photon singly
scattered\rq), the two diagrams Fig.~\ref{kerrc}(e) and (f) become identical.
In this case, we obtain a factor $2$, whereas the sum of diagram (e)
plus diagram (f)  yields $2+2=4$ in the case $n>1$.

Numerically, we solve the above integrals by a Monte-Carlo
method. Here, we proceed as follows:
for Eqs.~(\ref{ladderpol},\ref{crossedpol}),
at first position ${\bf r}$ and frequency $\omega_D$
of the inelastic scattering event are chosen randomly.
Starting from ${\bf r}$, three photons are launched, two with frequency
$\omega_L$ and one with frequency $\omega_D$. After each scattering event,
the length $r$ of the next propagation step is determined randomly according
to the distribution $P(r)=\exp(-r/\ell)/\ell$, whereas the
direction is chosen uniformly.
After all photons have left the medium,
the triple sum over $n$, $m$, and $l$ is performed, taking into
account the projection of the polarization vectors.
For the nonlinear propagation term, Eq.~(\ref{ellpol}), at first the
probe photon (path: $\bf u_1,\dots,{\bf u_n}$) is propagated, starting in the laser
mode
${\bf k_L},\epsilon_L$. Then, the pump photon is launched from
a randomly chosen position ${\bf v_m}$ on the path of the probe photon.
Finally, the projection of polarization vectors is performed separately for each
given path.

\end{document}